\pdfoutput=1

\documentclass[onecolumn]{emulateapj}

\usepackage{amsmath}
\usepackage{graphicx}
\usepackage{color}
\usepackage[colorlinks]{hyperref}
\hypersetup{
    colorlinks,	
    citecolor=blue,
}

\newcommand{\be}{\begin{eqnarray}}
\newcommand{\ee}{\end{eqnarray}}

\shorttitle{Tests of the Kerr hypothesis with GRS~1915+105}
\shortauthors{Zhang et al.}

\begin{document}

\title{Tests of the Kerr hypothesis with GRS~1915+105 using different {\sc relxill} flavors}

\author{Yuexin~Zhang\altaffilmark{1}, Askar~B.~Abdikamalov\altaffilmark{1}, Dimitry~Ayzenberg\altaffilmark{1}, Cosimo~Bambi\altaffilmark{1,\dag}, and Sourabh~Nampalliwar\altaffilmark{2}}

\altaffiltext{1}{Center for Field Theory and Particle Physics and Department of Physics, 
Fudan University, 200438 Shanghai, China. \email[\dag E-mail: ]{bambi@fudan.edu.cn}}
\altaffiltext{2}{Theoretical Astrophysics, Eberhard-Karls Universit\"at T\"ubingen, 72076 T\"ubingen, Germany}

\begin{abstract}
In a previous paper, we tried to test the Kerr nature of the stellar-mass black hole in GRS~1915+105 by analyzing \textsl{NuSTAR} data of 2012 with our reflection model {\sc relxill\_nk}. We found that the choice of the intensity profile of the reflection component is crucial and eventually we were not able to get any constraint on the spacetime metric around the black hole in GRS~1915+105. In the present paper, we study the same source with \textsl{Suzaku} data of 2007. We confirm that the intensity profile plays an important role, but now we find quite stringent constraints consistent with the Kerr hypothesis. The key differences with respect to our previous study are likely the lower disk temperature in the \textsl{Suzaku} observation and the higher energy resolution near the iron line of the \textsl{Suzaku} data. We also apply different {\sc relxill} flavors (different descriptions of the coronal spectrum and variable disk electron density) obtaining essentially the same results. We thus conclude that this choice is not very important for our tests of the Kerr hypothesis while the intensity profile does play an important role, and that with high quality data it is possible to measure both the spacetime metric and the intensity profile.
\end{abstract}


%%%%%%%%%%%%%%%%%%%%%%%%%%%%%%%

\section{Introduction}

In 4-dimensional Einstein's gravity, uncharged black holes are described by the Kerr solution~\citep{k1} and are completely specified by only two parameters, associated, respectively, with the mass $M$ and the spin angular momentum $J$ of the compact object~\citep{k2,k3}. The spacetime metric around astrophysical black holes formed from the complete gravitational collapse of stars or clouds is thought to be well approximated by the stationary, axisymmetric, and asymptotically-flat Kerr solution of Einstein's equations. Initial deviations from the Kerr geometry are expected to be quickly radiated away by the emission of gravitational waves after the black hole formation~\citep{kk1}. The presence of accretion disks or nearby stars around the black hole are normally negligible for the metric in the strong gravity region~\citep{kk2,kk3}. A possible initial non-vanishing electric charge is quickly almost neutralized because of the highly ionized host environment, and the residual equilibrium charge is very small and completely negligible for the spacetime geometry~\citep{kk4,kk5}. On the contrary, macroscopic deviations from the Kerr metric are possible in the presence of exotic matter~\citep{nk1,nk2}, quantum gravity effects~\citep{nk3,nk4,nk5}, as well as in a number of modified theories of gravity~\citep{nk6,nk7,nk8}.

Testing the Kerr nature of astrophysical black holes is becoming a hot topic nowadays, thanks to new observational facilities capable of probing the spacetime metric around these objects. Tests of the Kerr hypothesis involve either electromagnetic techniques~\citep{r1,r2} or gravitational waves~\citep{gw1,gw2,gw3,gw4}. Electromagnetic tests, strictly speaking, are sensitive to the motion of massive and massless particles in the strong gravity region and to couplings between the gravity and matter sectors. Gravitational wave tests are sensitive to the evolution of the spacetime metric in response to a variation of the mass/energy distribution and to the propagation of the gravitational wave signal. The two methods are thus complementary because they can test different sectors of the theory. For example, a new force inducing deviations from geodesic motion or variation of fundamental constants can naturally alter the electromagnetic spectrum of black holes, while the gravitational wave spectrum will be likely unchanged. A modified metric theory of gravity in which uncharged black holes are still described by the Kerr solution would predict the same electromagnetic spectrum as general relativity, because the spacetime metric is the same, but a different gravitational wave spectrum, because the field equations of the theory are different~\citep{gw11,gw12}.

There are many electromagnetic techniques proposed in literature to test the Kerr hypothesis~\citep{em1,em2,em3,em4,em5,em6,b12,em7,em8,em9}. Among all these methods, X-ray reflection spectroscopy seems to be the most promising for the present and near future, and surely the only one with observational constraints already published~\citep{noi1,noi2,noi3,noi4,noi5,noi6}. In the disk-corona model, thermal photons from the disk can inverse Compton scatter off free electrons in the so-called corona, which is some hot material ($\sim 100$~keV) in the strong gravity region. For example, the corona may be the base of the jet, the accretion flow plunging from the accretion disk to the black hole, the atmosphere above the accretion disk, etc. The Comptonized photons of the corona can illuminate the disk, producing a reflection component. The latter is characterized by fluorescent emission lines in the soft X-ray band (and the most prominent feature is often the iron K$\alpha$ line at 6.4-7~keV, depending on the ionization of iron atoms) and by the Compton hump peaked around 20-30~keV. The fluorescent emission lines are very narrow in the rest-frame of the gas, while they appear broadened and skewed far from the source as the results of relativistic effects occurring in the strong gravity region. X-ray reflection spectroscopy refers to the analysis of such a reflection component~\citep{ref1,ref2,ref3} and is potentially a powerful tool for investigating the strong gravity region of black holes.

{\sc relxill\_nk}~\citep{noim1,noim2} is an extension of the {\sc relxill} package~\citep{rx1,rx2,rx3} to non-Kerr backgrounds. The model describes the reflection spectrum of a Novikov-Thorne disk in a parametric black hole spacetime. The model parameters are the same as in the {\sc relxill} package together with some ``deformation parameters'' specifically introduced to deform the Kerr metric and quantify possible non-Kerr features. If all deformation parameters vanish, we exactly recover the Kerr background. From the comparison of the theoretical predictions of {\sc relxill\_nk} with observational data of accreting black holes with a strong reflection spectrum, it is possible to estimate the values of the deformation parameters and thus test the Kerr hypothesis.

In \citet{yuexin}, we applied {\sc relxill\_nk} to a 60~ks \textsl{NuSTAR} observation in 2012 of GRS~1915+105. This is quite a bright low mass X-ray binary with a stellar-mass black hole. In general, GRS~1915+105 is a highly variable source, but we showed that it was quite stable during the \textsl{NuSTAR} observation in 2012. The accretion luminosity of the black hole was around 20\% of its Eddington limit, so the thin disk model of {\sc relxill\_nk} was thought to be appropriate. Our results show instead that it is very complicated to test the Kerr metric of the stellar-mass black hole in GRS~1915+105 with the \textsl{NuSTAR} observation in 2012. Depending on the choice of the intensity profile, whether a simple power-law, a broken power-law with outer emissivity index fixed to 3, or a broken power-law with both emissivity indices free, we found different results, which may either confirm the Kerr metric or require deviations from the Kerr geometry. A similar dependence on the choice of the intensity profile is clearly what we do not want to have in a test of general relativity. Moreover, we found that the uncertainties on the deformation parameters were large in comparison with other measurements. We met similar problems with the analysis reported in~\citet{honghui} of a \textsl{NuSTAR} observation of another stellar-mass black hole, Cygnus~X-1. On the contrary, our studies of supermassive black holes seem to provide much more stringent constraints in agreement with the Kerr hypothesis and without any particular dependence on the choice of the adopted intensity profile, see~\citet{noi6} for a review of all results. So, it seems that supermassive black holes are more suitable than stellar-mass black holes for our tests of the Kerr metric.

In the present paper, we try to test the Kerr nature of the stellar-mass black hole in GRS~1915+105 by analyzing a \textsl{Suzaku} observation of 2007, previously studied in \citet{blum}. Here we also explore the impact of different {\sc relxill} flavors and we fit the data with {\sc relxill\_nk} (default model), {\sc relxillCp\_nk} (nthcomp Comptonization for the coronal spectrum), and {\sc relxillD\_nk} (variable disk electron density); see \citet{noim2} for more details on the specific flavors. We confirm that the choice of how to model the intensity profile of the reflection spectrum is crucial, but our conclusions are different. When we model the intensity profile of the reflection spectrum with a simple power-law, we do not recover the Kerr solution. When we employ a broken power-law with the two emissivity indices free in the fit, we find a better fit and we recover the Kerr metric. The result is not very sensitive to the exact {\sc relxill} flavor. This makes sense to us, because it says that the intensity profile is an important ingredient to properly model the reflection spectrum. A too simple intensity profile, like a simple power-law, is not enough to model high quality data like those of \textsl{Suzaku} of GRS~1915+105. When we employ a broken power-law with the emissivity indices free, the fit can both constrain the intensity profile and test the Kerr metric. We argue that there are two important differences between the \textsl{NuSTAR} observation analyzed in~\citet{yuexin} and the \textsl{Suzaku} one of the present work: $i)$ tests of the Kerr metric benefit from high energy resolution near the iron line, and $ii)$ in the \textsl{Suzaku} data we do not see any thermal component of the disk, while we see it in the \textsl{NuSTAR} data, suggesting that the temperature of the disk was much lower during the \textsl{Suzaku} observation. This is a relevant point because {\sc relxill\_nk} uses {\sc xillver}, in which the non-relativistic reflection spectrum is calculated assuming a cold disk (which indeed makes the model more suitable to study supermassive black holes rather than the stellar-mass ones).

The paper is organized as follows. In Section~\ref{s-obs}, we present the observation and how we reduced the data. In Section~\ref{s-ana}, we show the best-fit values and the constraints on the deformation parameters for all the models considered. In Section~\ref{s-dis}, we discuss the results. The parametric black hole metric employed in our study is reported in Appendix~\ref{a-metric}.

%%%%%%%%%%%%%%%%%%%%%%%%%%%%%%%

\section{Observation and data reduction}\label{s-obs}

GRS~1915+105 is a low mass X-ray binary with quite peculiar properties. In particular, it is a persistent X-ray source since its last outburst in 1992. \textsl{Suzaku} observed GRS~1915+105 on 2007 May 7 (obs. ID~402071010) for approximately 117~ks. In our analysis, we used the data from the XIS1 and HXD/PIN instruments only. Two other XIS units were turned off to preserve telemetry, and the fourth unit was run in a special timing mode. We employed Xspec~v12.10.0~\citep{xspec}.

We processed unfiltered event files of the XIS1 following the Suzaku Data Reduction ABC Guide with \texttt{aepipeline} to create a clean event file ($3\times 3$ mode and $5\times 5$ mode data), using XIS CALDB version 20160616. The source region was selected by an annulus region of inner radius $78''$ and outer radius $208''$ because photons piled up severely in the center of the detector~\citep{blum}. The background region was selected by an annulus region of inner radius $208''$ and outer radius $278''$. We removed the extracted region areas that did not land on the XIS detector manually. Unbinned source and background spectra were extracted with \texttt{xselect}, ensuring the cutoff-rigidity was set to $>6$~GeV to account for proper non X-ray background (NXB) subtraction. The XIS redistribution matrix file (RMF) and ancillary response file (ARF) were created respectively using the tools \texttt{xisrmfgen} and \texttt{xissimarfgen} available in the HEASOFT version 6.24 data reduction package. After all efficiencies and screening, a net exposure time of 28.94~ks for the XIS1 camera (in the $3\times 3$ editing mode) was achieved. We grouped the data to a minimum of 25 counts per bin using \texttt{grppha}. In our analysis, we used the 2.3-10 keV energy band in order to avoid calibration problems near the Si~K edge and because there are few photons at low energy due to absorption from high column density, which could have negatively influenced our estimate of the iron K$\alpha$ emission line.

\begin{figure}[t]
\begin{center}
\vspace{0.3cm}
\includegraphics[width=0.49\textwidth]{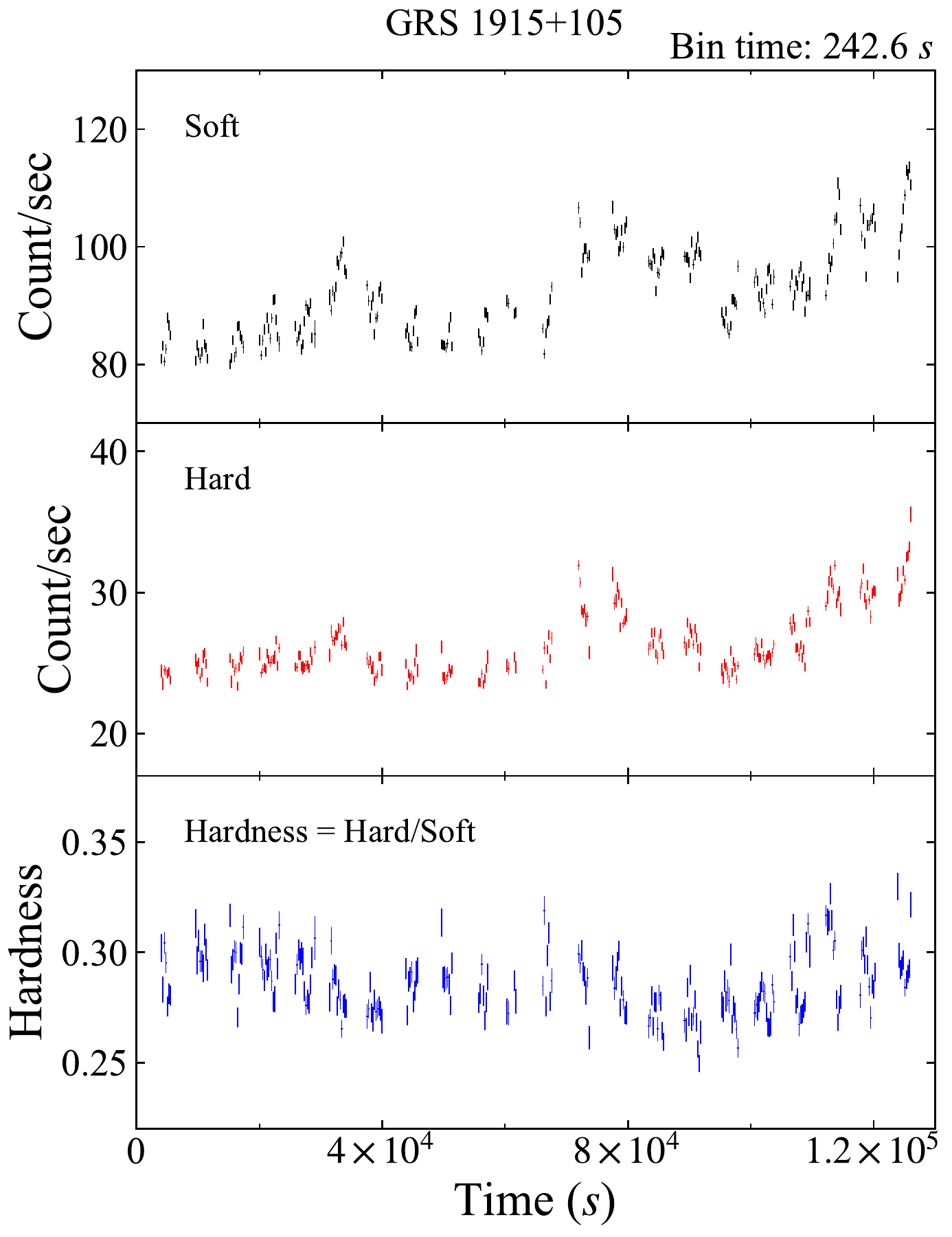}
\end{center}
%\vspace{-1.0cm}
\caption{Light curves in the soft band (0.2-12~keV, XIS1 data, upper panel) and hard band (10-70~keV, HXD/PIN data, central panel) and temporal evolution of the hardness of the spectrum (bottom panel) of GRS~1915+105 on 2007 May 7. \label{f-lightcurve}}
\vspace{1.0cm}
\end{figure}

HXD/PIN data were reduced similarly, employing \texttt{aepipeline} and then \texttt{hxdpinxbpi} using the latest CALDB version 20110915. After all efficiencies and screening, a net exposure time of 53.00~ks for the HXD/PIN was achieved. In our analysis, we used the 12.0--55.0 keV energy band following~\citet{blum}. We used a cross-normalization constant floating between XIS1 data and HXD/PIN data\footnote{Note that the recommended cross calibration constant 1.16 between XIS and HXD/PIN instruments applies to simulations and sources like AGNs. In the case of bright sources, like GRS~1915+105, the XIS instruments are affected by pile-up while HXD/PIN is not. Removing the central part in the XIS image has an impact of the measured flux of the XIS instrument and therefore we have to fit the cross-normalization constant.}.

In general, GRS~1915+105 is a highly variable source. However, the hardness of the source was quite stable during the \textsl{Suzaku} observation of 2007, and we did not need to take its variability into account in the spectral analysis. The light curve is shown in Fig.~\ref{f-lightcurve}.

%%%%%%%%%%%%%%%%%%%%%%%%%%%%%%%

\begin{table}
\centering
\caption{ \label{t-kerr}}
\begin{tabular}{lcc|cc|cc}
\hline\hline
 & \multicolumn{2}{c}{\sc relxill\_nk} & \multicolumn{2}{c}{\sc relxillCp\_nk} & \multicolumn{2}{c}{\sc relxillD\_nk} \\
\hline
{\sc tbabs} &&&&&& \\
$N_{\rm H} / 10^{22}$ cm$^{-2}$ & $7.92^{+0.05}_{-0.04}$ & $8.86^{+0.17}_{-0.06}$ & $7.90^{+0.10}_{-0.06}$ & $9.33^{+0.07}_{-0.06}$ & $8.28^{+0.05}_{-0.07}$ & $9.19^{+0.06}_{-0.04}$ \\
\hline
{\sc relxill\_nk} &&&&&& \\
$q_{\rm in}$ & $10.0_{-0.3}$ & $10.0_{-1.6}$ & $10.0_{-0.4}$ & $10.0_{-0.2}$ & $1.88^{+0.08}_{-0.07}$ & $7.36^{+0.07}_{-0.42}$ \\
$q_{\rm out}$ & $=q_{\rm in}$ & $0.0^{+0.3}$ & $=q_{\rm in}$ & $0.0^{+0.3}$ & $=q_{\rm in}$ & $0.0^{+0.2}$ \\
$R_{\rm br}$~$[M]$ & $...$ & $6.1^{+0.4}_{-0.2}$ & $...$ & $6.0^{+0.3}_{-0.2}$ & $...$ & $10.4^{+0.2}_{-0.2}$ \\
$i$ [deg] & $65^{+1}_{-1}$ & $74.6^{+0.7}_{-1.0}$ & $62^{+1}_{-2}$ & $74.7^{+0.6}_{-1.2}$ & $80_{-2}$ & $69.2^{+0.7}_{-0.5}$ \\
$a_*$ & $0.960^{+0.007}_{-0.008}$ & $0.991^{+0.001}_{-0.001}$ & $0.94^{+0.01}_{-0.02}$ & $0.992^{+0.001}_{-0.001}$ & $0.90^{+0.07}_{-0.10}$ & $0.998_{-0.002}$ \\
$\alpha_{13}$ & $0^\star$ & $0^\star$ & $0^\star$ & $0^\star$ & $0^\star$ & $0^\star$ \\
$\alpha_{22}$ & $0^\star$ & $0^\star$ & $0^\star$ & $0^\star$ & $0^\star$ & $0^\star$ \\
$\epsilon_{3}$ & $0^\star$ & $0^\star$ & $0^\star$ & $0^\star$ & $0^\star$ & $0^\star$ \\
$\log\xi$ & $2.94^{+0.02}_{-0.04}$ & $2.54^{+0.10}_{-0.04}$ & $3.07^{+0.02}_{-0.02}$ & $2.49^{+0.04}_{-0.03}$ & $0.66^{+0.08}_{-0.26}$ & $2.50^{+0.01}_{-0.18}$ \\
$A_{\rm Fe}$ & $0.56^{+0.05}_{-0.02}$ & $0.66^{+0.05}_{-0.02}$ & $0.50^{+0.02}_{}$ & $0.57^{+0.04}_{-0.04}$ & $0.63^{+0.07}_{-0.08}$ & $0.70^{+0.03}_{-0.04}$ \\
$\Gamma$ & $2.20^{+0.01}_{-0.02}$ & $2.44^{+0.06}_{-0.02}$ & $2.203^{+0.007}_{-0.008}$ & $2.527^{+0.013}_{-0.004}$ & $2.74^{+0.02}_{-0.03}$ & $2.587^{+0.008}_{-0.006}$ \\
$E_{\rm cut}$($kT_{\rm e}$) [keV] & $74^{+2}_{-2}$ & $110^{+5}_{-8}$ & $26^{+2}_{-2}$ & $200^{}_{-43}$ & $300^\star$ & $300^\star$ \\
$\log{\rm N}$ [cm$^{-3}$] & $...$ & $...$ & $...$ & $...$ & $17.6^{+0.2}_{-0.1}$ & $16.62^{+0.08}_{-0.04}$ \\
$R_{\rm f}$ & $0.59^{+0.05}_{-0.03}$ & $0.71^{+0.05}_{-0.05}$ & $0.53^{+0.05}_{-0.03}$ & $1.01^{+0.05}_{-0.04}$ & $0.7^{+0.2}_{-0.1}$ & $1.22^{+0.04}_{-0.11}$\\
norm & $0.0440_{-0.0003}^{+0.0007}$ & $0.068_{-0.001}^{+0.005}$ & $0.0387_{-0.0002}^{+0.0020}$ & $0.0648_{-0.0005}^{+0.0002}$ & $0.146_{-0.006}^{+0.006}$ & $0.0782_{-0.0002}^{+0.0040}$ \\
\hline
$\chi^2/\nu$ & $\quad 2347.37/2184 \quad$ & $\quad 2258.59/2182 \quad$ &  $\quad 2400.32/2211 \quad$ & $\quad 2289.62/2209 \quad$ & $\quad 2328.45/2211 \quad$ & $\quad 2289.65/2209 \quad$\\
& =1.07481 & =1.03510 & =1.08562 & =1.03649 & =1.05312 & =1.03651 \\
\hline\hline
\end{tabular}
\vspace{0.2cm}
\tablenotetext{0}{Best-fit values for the Kerr model ($\alpha_{13} = \alpha_{22} = \epsilon_3 = 0$) with {\sc relxill\_nk},  {\sc relxillCp\_nk}, and {\sc relxillD\_nk}. For every flavor, we have two models for the emissivity profile: simple power-law (left column) and broken power-law (right column). The reported uncertainties correspond to the 90\% confidence level for one relevant parameter. $^\star$ indicates that the value is frozen in the fit. $q_{\rm in}$ and $q_{\rm out}$ are allowed to range between 0 and 10, and they often get stuck at the boundaries. $i$ is allowed to vary from 3$^\circ$ to 80$^\circ$. $a_*$ is allowed to vary from $-0.998$ to $0.998$. $A_{\rm Fe}$ is allowed to range from 0.5 to 10. $kT_{\rm e}$ in {\sc relxillCp\_nk} can vary from 1~keV to 200~keV. $\log{\rm N}$ in {\sc relxillD\_nk} is allowed to range from 15 to 19. When the lower/upper uncertainty is not reported, the 90\% confidence level reaches the boundary (or the best-fit is at the boundary).}
\vspace{0.2cm}
%\end{table}
%\begin{table}
\centering
\caption{ \label{t-a13}}
\begin{tabular}{lcccccc}
\hline\hline
 & \multicolumn{2}{c}{\sc relxill\_nk} & \multicolumn{2}{c}{\sc relxillCp\_nk} & \multicolumn{2}{c}{\sc relxillD\_nk} \\
\hline
{\sc tbabs} &&&&&& \\
$N_{\rm H} / 10^{22}$ cm$^{-2}$ & $8.94^{+0.08}_{-0.22}$ & $8.85^{+0.15}_{-0.08}$ & $8.95^{+0.05}_{-0.07}$ & $9.32^{+0.07}_{-0.09}$ & $8.29^{+0.03}_{-0.05}$ & $9.29^{+0.07}_{-0.10}$ \\
\hline
{\sc relxill\_nk} &&&&&& \\
$q_{\rm in}$ & $5.0^{+0.1}_{-0.2}$ & $10.0_{-0.9}$ & $5.0^{+0.2}_{-0.2}$ & $10.0_{-4.8}$ & $1.91^{+0.08}_{-0.05}$ & $6.49^{+0.07}_{-0.16}$ \\
$q_{\rm out}$ & $=q_{\rm in}$ & $0.0^{+0.4}$ & $=q_{\rm in}$ & $0.0^{+0.2}$ & $=q_{\rm in}$ & $0.00^{+0.14}$ \\
$R_{\rm br}$~$[M]$ & $...$ & $5.9^{+0.9}_{-2.5}$ & $...$ & $5.3^{+1.4}_{-0.4}$ & $...$ & $12.3^{+4.5}_{-0.4}$ \\
$i$ [deg] & $67.1^{+0.4}_{-0.5}$ & $75^{+2}_{-2}$ & $66.9^{+0.4}_{-0.5}$ & $76.7^{+0.7}_{-2.8}$ & $80^{}_{-1}$ & $69.2^{+0.7}_{-0.4}$ \\
$a_*$ & $0.998_{-0.001}$ & $0.992_{-0.004}$ & $0.998_{-0.002}$ & $0.996_{-0.006}$ & $0.998_{-0.197}$ & $0.998_{-0.002}$ \\
$\alpha_{13}$ & $-0.44^{+0.04}_{-0.01}$ & $-0.05^{+0.05}_{-0.18}$ & $-0.44^{+0.04}_{-0.01}$ & $-0.17^{+0.20}_{-0.02}$ & $0.6^{+1.1}_{-0.8}$ & $-0.18^{+0.12}_{-0.03}$ \\
$\alpha_{22}$ & $0^\star$ & $0^\star$ & $0^\star$ & $0^\star$ & $0^\star$ & $0^\star$ \\
$\epsilon_{3}$ & $0^\star$ & $0^\star$ & $0^\star$ & $0^\star$ & $0^\star$ & $0^\star$ \\
$\log\xi$ & $2.74^{+0.06}_{-0.03}$ & $2.54^{+0.09}_{-0.07}$ & $2.74^{+0.03}_{-0.03}$ & $2.49^{+0.05}_{-0.04}$ & $0.65^{+0.07}_{-0.24}$ & $2.54^{+0.04}_{-0.06}$ \\
$A_{\rm Fe}$ & $0.50^{+0.04}_{}$ & $0.67^{+0.08}_{-0.06}$ & $0.50^{+0.02}_{}$ & $0.58^{+0.05}_{-0.04}$ & $0.66^{+0.06}_{-0.03}$ & $0.68^{+0.02}_{-0.03}$ \\
$\Gamma$ & $2.40^{+0.02}_{-0.03}$ & $2.44^{+0.07}_{-0.05}$ & $2.40^{+0.01}_{-0.01}$ & $2.53^{+0.02}_{-0.03}$ & $2.75^{+0.02}_{-0.03}$ & $2.585^{+0.014}_{-0.003}$ \\
$E_{\rm cut}$($kT_{\rm e}$) [keV] & $116^{+7}_{-18}$ & $110^{+13}_{-8}$ & $68^{+7}_{-9}$ & $200^{}_{-23}$ & $300^\star$ & $300^\star$ \\
$\log{\rm N}$ [cm$^{-3}$] & $...$ & $...$ & $...$ & $...$ & $17.5^{+0.2}_{-0.1}$ & $16.65^{+0.26}_{-0.08}$ \\
$R_{\rm f}$ & $0.78^{+0.05}_{-0.04}$ & $0.71^{+0.07}_{-0.06}$ & $0.86^{+0.04}_{-0.05}$ & $1.01^{+0.09}_{-0.10}$ & $0.74^{+0.18}_{-0.17}$ & $1.21^{+0.05}_{-0.06}$\\
norm & $0.062_{-0.007}^{+0.004}$ & $0.068_{-0.004}^{+0.005}$ & $0.054_{-0.003}^{+0.003}$ & $0.065_{-0.006}^{+0.003}$ & $0.149_{-0.005}^{+0.005}$ & $0.078_{-0.004}^{+0.002}$ \\
\hline
$\chi^2/\nu$ & $\quad 2285.11/2183 \quad$ & $\quad 2258.34/2181 \quad$ &  $\quad 2337.42/2210 \quad$ & $\quad 2288.72/2208 \quad$ & $\quad 2327.05/2210 \quad$ & $\quad 2288.68/2208 \quad$\\
& =1.04678 & =1.03546 & =1.05765 & =1.03656 & =1.05296 & =1.03654 \\
\hline\hline
\end{tabular}
\vspace{0.2cm}
\tablenotetext{0}{Best-fit values for the Johannsen model with free $\alpha_{13}$ and with {\sc relxill\_nk},  {\sc relxillCp\_nk}, and {\sc relxillD\_nk}. For every flavor, we have two models for the emissivity profile: simple power-law (left column) and broken power-law (right column). The reported uncertainties correspond to the 90\% confidence level for one relevant parameter. $^\star$ indicates that the value is frozen in the fit. $q_{\rm in}$ and $q_{\rm out}$ are allowed to range between 0 and 10, and they often get stuck at the boundaries. $i$ is allowed to vary from 3$^\circ$ to 80$^\circ$. $a_*$ is allowed to vary from $-0.998$ to $0.998$. $A_{\rm Fe}$ is allowed to range from 0.5 to 10. $kT_{\rm e}$ in {\sc relxillCp\_nk} can vary from 1~keV to 200~keV. $\log{\rm N}$ in {\sc relxillD\_nk} is allowed to range from 15 to 19. When the lower/upper uncertainty is not reported, the 90\% confidence level reaches the boundary (or the best-fit is at the boundary).}
\end{table}

\begin{table}
\centering
\caption{ \label{t-a22}}
\begin{tabular}{lcccccc}
\hline\hline
 & \multicolumn{2}{c}{\sc relxill\_nk} & \multicolumn{2}{c}{\sc relxillCp\_nk} & \multicolumn{2}{c}{\sc relxillD\_nk} \\
\hline
{\sc tbabs} &&&&&& \\
$N_{\rm H} / 10^{22}$ cm$^{-2}$ & $8.8^{+0.2}_{-0.5}$ & $8.86^{+0.09}_{-0.05}$ & $8.90^{+0.05}_{-0.04}$ & $9.31^{+0.07}_{-0.03}$ & $8.27^{+0.03}_{-0.05}$ & $9.3^{+0.1}_{-0.1}$ \\
\hline
{\sc relxill\_nk} &&&&&& \\
$q_{\rm in}$ & $5.0^{+0.2}_{-0.3}$ & $10.0_{-0.3}$ & $4.9^{+0.2}_{-0.1}$ & $10.0_{-0.3}$ & $1.88^{+0.07}_{-0.06}$ & $6.9^{+0.2}_{-0.5}$ \\
$q_{\rm out}$ & $=q_{\rm in}$ & $0.0^{+0.4}$ & $=q_{\rm in}$ & $0.0^{+0.2}$ & $=q_{\rm in}$ & $0.0^{+0.4}$ \\
$R_{\rm br}$~$[M]$ & $...$ & $6.1^{+1.4}_{-0.4}$ & $...$ & $5.8^{+0.5}_{-0.7}$ & $...$ & $11^{+1}_{-5}$ \\
$i$ [deg] & $64.5^{+0.8}_{-0.8}$ & $74.7^{+0.6}_{-1.4}$ & $63.9^{+0.9}_{-0.3}$ & $75.0^{+0.2}_{-0.4}$ & $80.0^{}_{-1.3}$ & $69.8^{+1.3}_{-0.8}$ \\
$a_*$ & $0.998_{-0.001}$ & $0.991^{+0.004}_{-0.003}$ & $0.998_{-0.001}$ & $0.990^{+0.003}_{-0.001}$ & $0.993_{-0.471}$ & $0.998_{-0.002}$ \\
$\alpha_{13}$ & $0^\star$ & $0^\star$ & $0^\star$ & $0^\star$ & $0^\star$ & $0^\star$ \\
$\alpha_{22}$ & $0.59^{+0.02}_{-0.03}$ & $0.003^{+0.100}_{-0.013}$ & $0.62^{+0.01}_{-0.05}$ & $0.10^{+0.03}_{-0.11}$ & $-0.4^{+0.6}_{-0.2}$ & $0.19^{+0.06}_{-0.18}$ \\
$\epsilon_{3}$ & $0^\star$ & $0^\star$ & $0^\star$ & $0^\star$ & $0^\star$ & $0^\star$ \\
$\log\xi$ & $2.77^{+0.11}_{-0.03}$ & $2.53^{+0.06}_{-0.04}$ & $2.74^{+0.03}_{-0.03}$ & $2.51^{+0.04}_{-0.02}$ & $0.67^{+0.41}_{-0.05}$ & $2.42^{+0.03}_{-0.03}$ \\
$A_{\rm Fe}$ & $0.50^{+0.03}_{}$ & $0.67^{+0.05}_{-0.05}$ & $0.50^{+0.02}_{}$ & $0.57^{+0.03}_{-0.02}$ & $0.63^{+0.10}_{-0.03}$ & $0.79^{+0.04}_{-0.04}$ \\
$\Gamma$ & $2.38^{+0.03}_{-0.10}$ & $2.44^{+0.06}_{-0.01}$ & $2.400^{+0.010}_{-0.009}$ & $2.52^{+0.01}_{-0.01}$ & $2.739^{+0.007}_{-0.031}$ & $2.581^{+0.010}_{-0.004}$ \\
$E_{\rm cut}$($kT_{\rm e}$) [keV] & $109^{+15}_{-14}$ & $110^{+5}_{-7}$ & $68^{+7}_{-7}$ & $200^{}_{-43}$ & $300^\star$ & $300^\star$ \\
$\log{\rm N}$ [cm$^{-3}$] & $...$ & $...$ & $...$ & $...$ & $17.54^{+0.19}_{-0.02}$ & $17.60^{+0.11}_{-0.07}$ \\
$R_{\rm f}$ & $0.76^{+0.08}_{-0.10}$ & $0.70^{+0.08}_{-0.10}$ & $0.84^{+0.04}_{-0.04}$ & $0.99^{+0.02}_{-0.14}$ & $0.73^{+0.05}_{-0.16}$ & $1.2^{+0.1}_{-0.1}$\\
norm & $0.059_{-0.011}^{+0.009}$ & $0.068_{-0.003}^{+0.005}$ & $0.055_{-0.003}^{+0.003}$ & $0.064_{-0.002}^{+0.003}$ & $0.145_{-0.006}^{+0.002}$ & $0.075_{-0.006}^{+0.002}$ \\
\hline
$\chi^2/\nu$ & $\quad 2294.32/2183 \quad$ & $\quad 2258.86/2181 \quad$ &  $\quad 2345.30/2210 \quad$ & $\quad 2288.24/2208 \quad$ & $\quad 2328.38/2210 \quad$ & $\quad 2284.61/2208 \quad$\\
& =1.05099 & =1.03570 & =1.06122 & =1.03634 & =1.05357 & =1.03470 \\
\hline\hline
\end{tabular}
\vspace{0.2cm}
\tablenotetext{0}{Best-fit values for the Johannsen model with free $\alpha_{22}$ and with {\sc relxill\_nk},  {\sc relxillCp\_nk}, and {\sc relxillD\_nk}. For every flavor, we have two models for the emissivity profile: simple power-law (left column) and broken power-law (right column). The reported uncertainties correspond to the 90\% confidence level for one relevant parameter. $^\star$ indicates that the value is frozen in the fit. $q_{\rm in}$ and $q_{\rm out}$ are allowed to range between 0 and 10, and they often get stuck at the boundaries. $i$ is allowed to vary from 3$^\circ$ to 80$^\circ$. $a_*$ is allowed to vary from $-0.998$ to $0.998$. $A_{\rm Fe}$ is allowed to range from 0.5 to 10. $kT_{\rm e}$ in {\sc relxillCp\_nk} can vary from 1~keV to 200~keV. $\log{\rm N}$ in {\sc relxillD\_nk} is allowed to range from 15 to 19. When the lower/upper uncertainty is not reported, the 90\% confidence level reaches the boundary (or the best-fit is at the boundary).}
\end{table}

\begin{table}
\centering
\caption{ \label{t-e3}}
\begin{tabular}{lcccccc}
\hline\hline
 & \multicolumn{2}{c}{\sc relxill\_nk} & \multicolumn{2}{c}{\sc relxillCp\_nk} & \multicolumn{2}{c}{\sc relxillD\_nk} \\
\hline
{\sc tbabs} &&&&&& \\
$N_{\rm H} / 10^{22}$ cm$^{-2}$ & $7.90^{+0.04}_{-0.04}$ & $8.86^{+0.09}_{-0.07}$ & $7.96^{+0.02}_{-0.02}$ & $9.33^{+0.08}_{-0.06}$ & $8.29^{+0.08}_{-0.07}$ & $9.19^{+0.04}_{-0.06}$ \\
\hline
{\sc relxill\_nk} &&&&&& \\
$q_{\rm in}$ & $10.0_{-0.3}$ & $10.0_{-0.6}$ & $10.0_{-0.3}$ & $10.0_{-0.4}$ & $1.91^{+0.09}_{-0.05}$ & $7.36^{+0.06}_{-0.37}$ \\
$q_{\rm out}$ & $=q_{\rm in}$ & $0.0^{+0.5}$ & $=q_{\rm in}$ & $0.0^{+0.3}$ & $=q_{\rm in}$ & $0.0^{+0.2}$ \\
$R_{\rm br}$~$[M]$ & $...$ & $6.2^{+2.3}_{-0.8}$ & $...$ & $6.0^{+0.3}_{-0.2}$ & $...$ & $10.4^{+1.0}_{-0.4}$ \\
$i$ [deg] & $67.2^{+0.6}_{-0.8}$ & $74.6^{+0.7}_{-1.4}$ & $64.8^{+0.2}_{-0.7}$ & $74.7^{+0.6}_{-1.2}$ & $79.0^{}_{-0.5}$ & $69.2^{+0.5}_{-0.5}$ \\
$a_*$ & $0.991^{+0.004}_{-0.003}$ & $0.989^{+0.006}_{-0.003}$ & $0.986^{+0.006}_{-0.021}$ & $0.992^{+0.004}_{-0.005}$ & $0.85^{+0.09}_{-0.13}$ & $0.998_{-0.003}$ \\
$\alpha_{13}$ & $0^\star$ & $0^\star$ & $0^\star$ & $0^\star$ & $0^\star$ & $0^\star$ \\
$\alpha_{22}$ & $0^\star$ & $0^\star$ & $0^\star$ & $0^\star$ & $0^\star$ & $0^\star$ \\
$\epsilon_{3}$ & $1.0^{+0.1}_{-0.8}$ & $-0.19^{+0.35}_{-0.03}$ & $0.94^{+0.60}_{-0.30}$ & $0.00^{+0.30}_{-0.05}$ & $-3.1^{+3.3}_{-2.0}$ & $0.03^{+0.20}_{-1.78}$ \\
$\log\xi$ & $2.89^{+0.02}_{-0.07}$ & $2.57^{+0.08}_{-0.09}$ & $3.037^{+0.005}_{-0.015}$ & $2.49^{+0.04}_{-0.03}$ & $0.69^{+0.04}_{-0.26}$ & $2.50^{+0.02}_{-0.03}$ \\
$A_{\rm Fe}$ & $0.58^{+0.03}_{-0.03}$ & $0.68^{+0.05}_{-0.08}$ & $0.50^{+0.01}_{}$ & $0.57^{+0.04}_{-0.04}$ & $0.61^{+0.08}_{-0.04}$ & $0.70^{+0.03}_{-0.03}$ \\
$\Gamma$ & $2.203^{+0.007}_{-0.006}$ & $2.44^{+0.04}_{-0.05}$ & $2.22^{+0.01}_{-0.02}$ & $2.527^{+0.009}_{-0.024}$ & $2.74^{+0.02}_{-0.03}$ & $2.587^{+0.015}_{-0.005}$ \\
$E_{\rm cut}$($kT_{\rm e}$) [keV] & $72^{+1}_{-2}$ & $112^{+13}_{-7}$ & $26.9^{+0.5}_{-0.9}$ & $200^{}_{-45}$ & $300^\star$ & $300^\star$ \\
$\log{\rm N}$ [cm$^{-3}$] & $...$ & $...$ & $...$ & $...$ & $17.62^{+0.12}_{-0.04}$ & $16.62^{+0.08}_{-0.04}$ \\
$R_{\rm f}$ & $0.55^{+0.02}_{-0.02}$ & $0.74^{+0.06}_{-0.04}$ & $0.56^{+0.01}_{-0.07}$ & $1.01^{+0.07}_{-0.07}$ & $0.80^{+0.07}_{-0.18}$ & $1.22^{+0.05}_{-0.02}$\\
norm & $0.0452_{-0.0010}^{+0.0002}$ & $0.067_{-0.004}^{+0.006}$ & $0.0400_{-0.0003}^{+0.0013}$ & $0.065_{-0.003}^{+0.003}$ & $0.147_{-0.005}^{+0.006}$ & $0.078_{-0.004}^{+0.004}$ \\
\hline
$\chi^2/\nu$ & $\quad 2344.21/2183 \quad$ & $\quad 2258.26/2181 \quad$ &  $\quad 2395.96/2210 \quad$ & $\quad 2289.59/2208 \quad$ & $\quad 2326.82/2210 \quad$ & $\quad 2289.48/2208 \quad$\\
& =1.07385 & =1.03543 & =1.08414 & =1.03695 & =1.05286 & =1.03690 \\
\hline\hline
\end{tabular}
\vspace{0.2cm}
\tablenotetext{0}{Best-fit values for the Johannsen model with free $\epsilon_3$ and with {\sc relxill\_nk},  {\sc relxillCp\_nk}, and {\sc relxillD\_nk}. For every flavor, we have two models for the emissivity profile: simple power-law (left column) and broken power-law (right column). The reported uncertainties correspond to the 90\% confidence level for one relevant parameter. $^\star$ indicates that the value is frozen in the fit. $q_{\rm in}$ and $q_{\rm out}$ are allowed to range between 0 and 10, and they often get stuck at the boundaries. $i$ is allowed to vary from 3$^\circ$ to 80$^\circ$. $a_*$ is allowed to vary from $-0.998$ to $0.998$. $A_{\rm Fe}$ is allowed to range from 0.5 to 10. $kT_{\rm e}$ in {\sc relxillCp\_nk} can vary from 1~keV to 200~keV. $\log{\rm N}$ in {\sc relxillD\_nk} is allowed to range from 15 to 19. When the lower/upper uncertainty is not reported, the 90\% confidence level reaches the boundary (or the best-fit is at the boundary).}
\end{table}

\begin{figure*}[t]
\begin{center}
\vspace{0.3cm}
\includegraphics[width=0.45\textwidth]{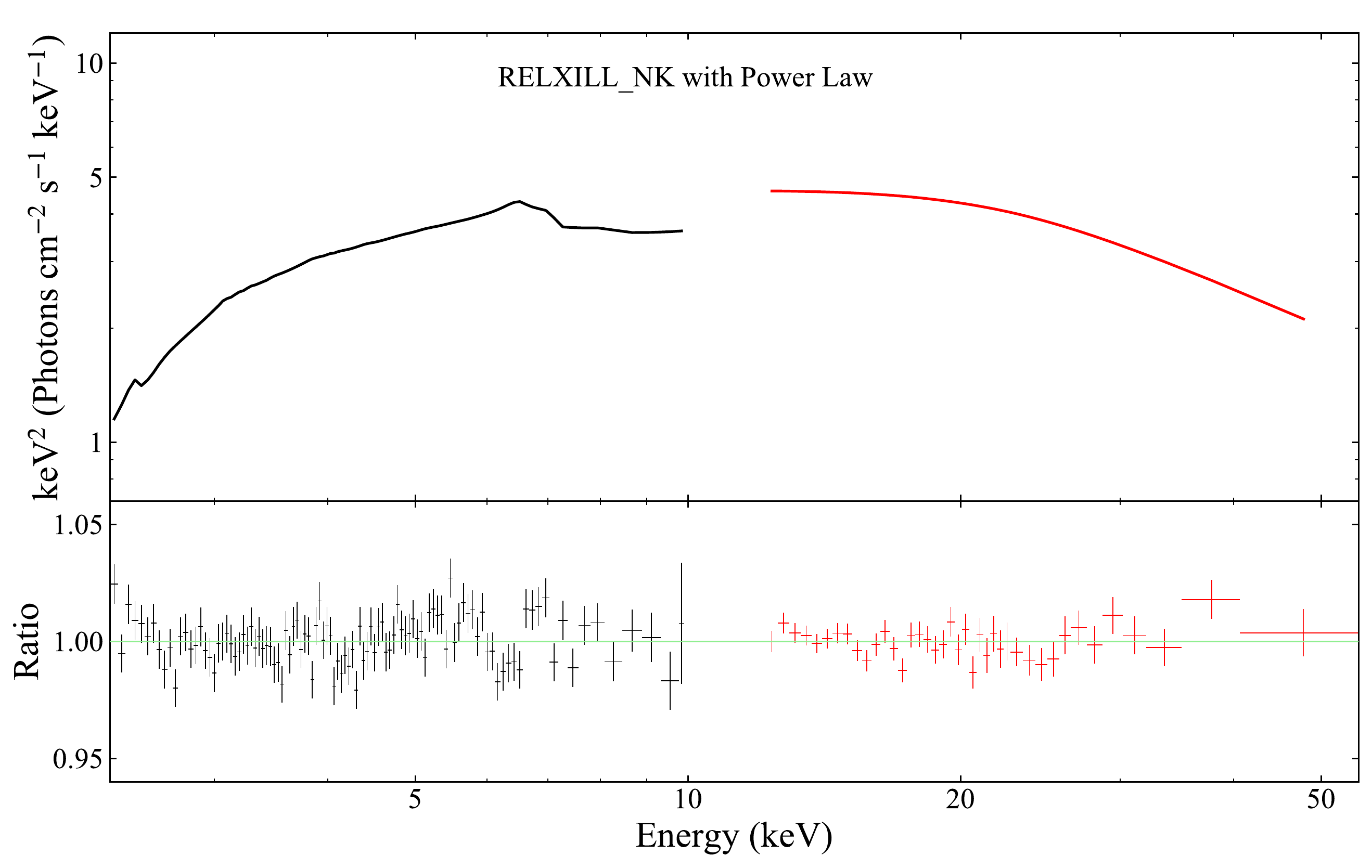}
\hspace{0.4cm}
\includegraphics[width=0.45\textwidth]{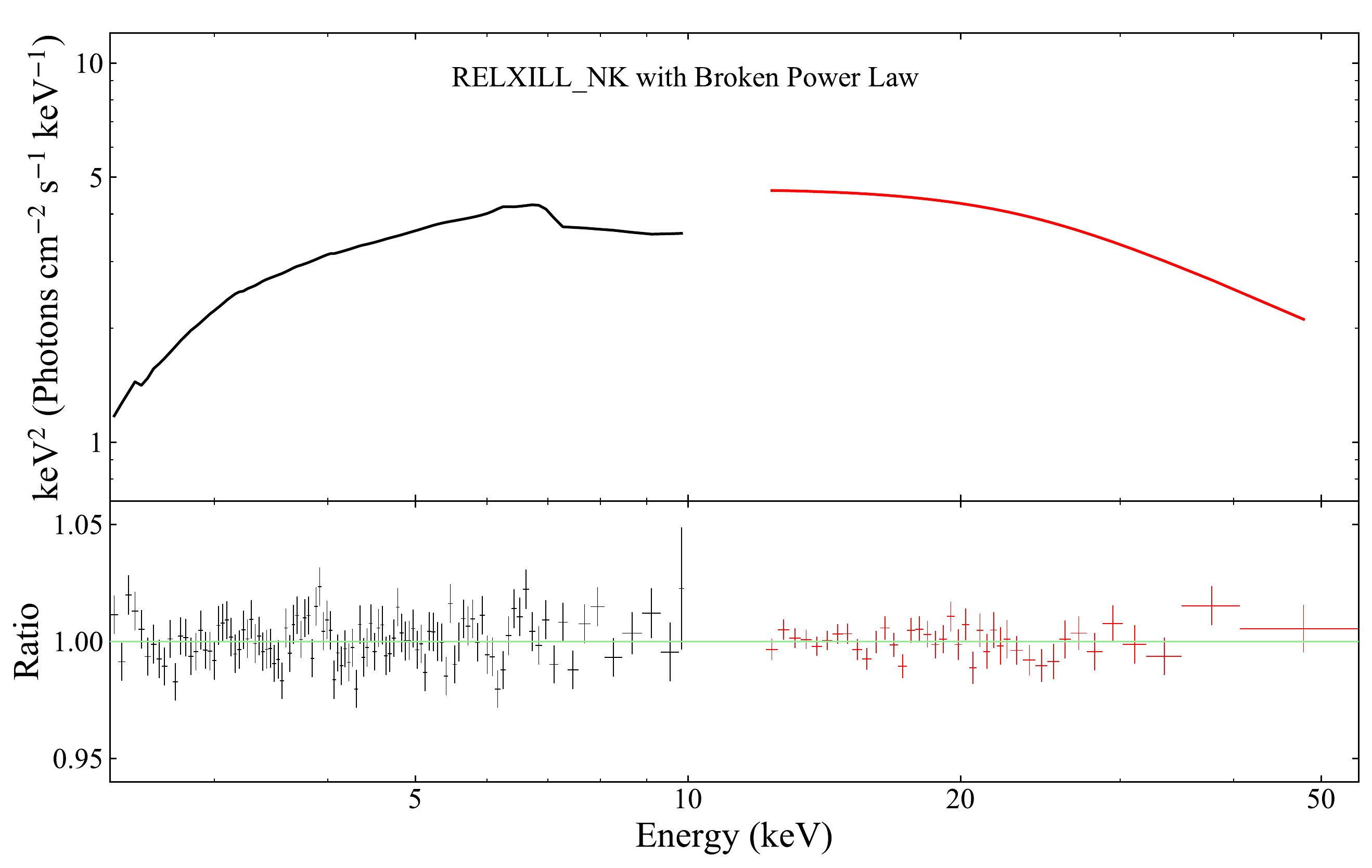} 
\vspace{0.2cm} \\
\includegraphics[width=0.45\textwidth]{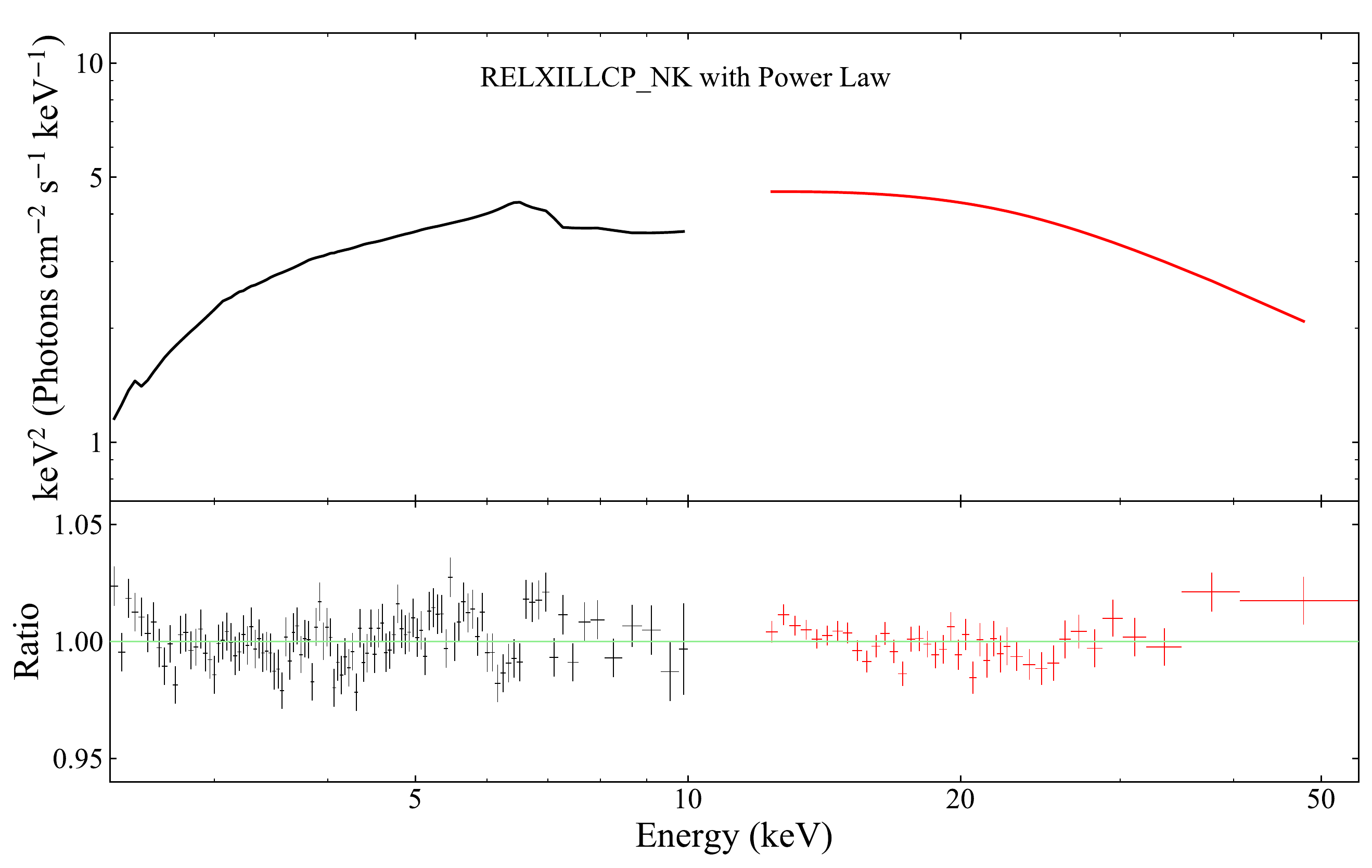}
\hspace{0.4cm}
\includegraphics[width=0.45\textwidth]{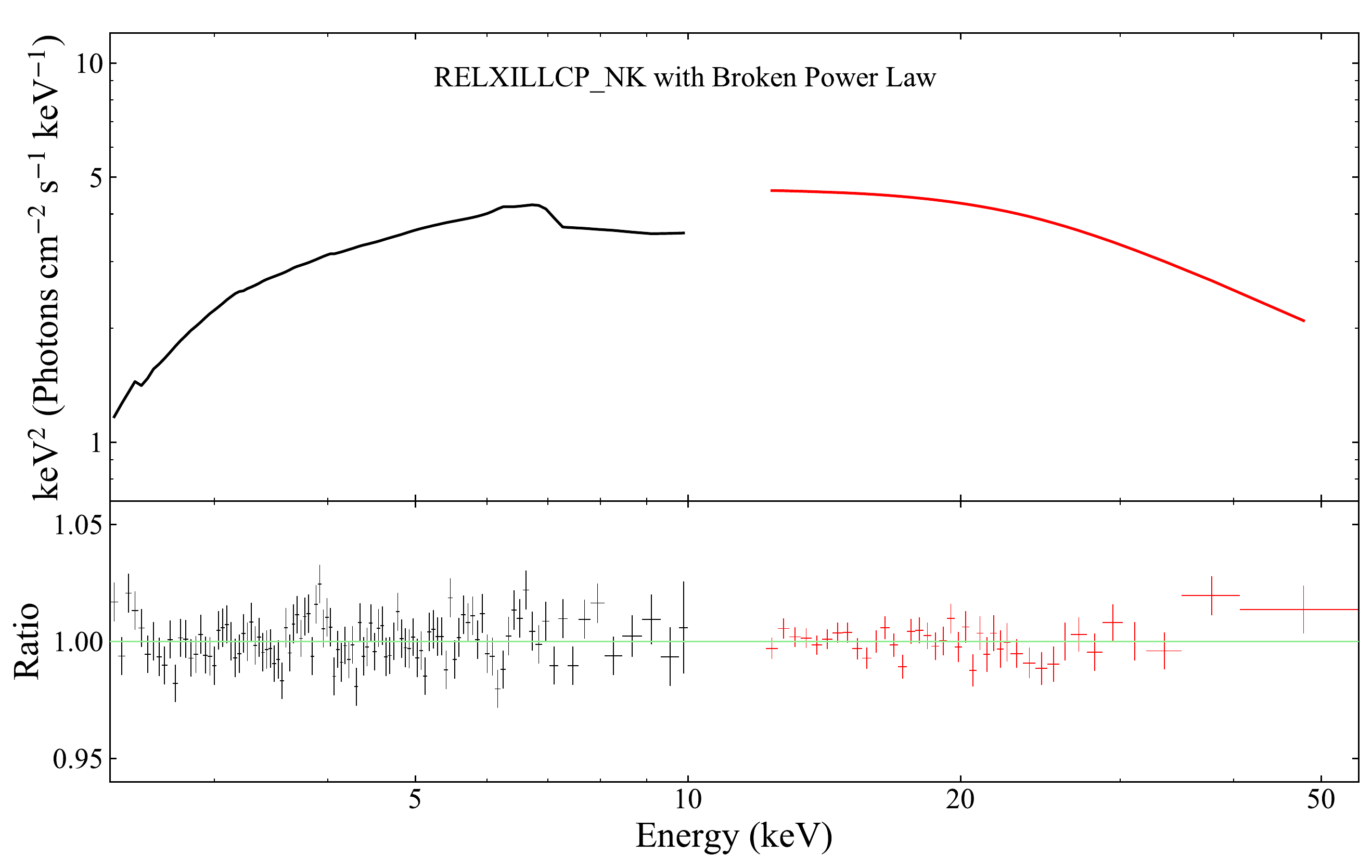} 
\vspace{0.2cm} \\
\includegraphics[width=0.45\textwidth]{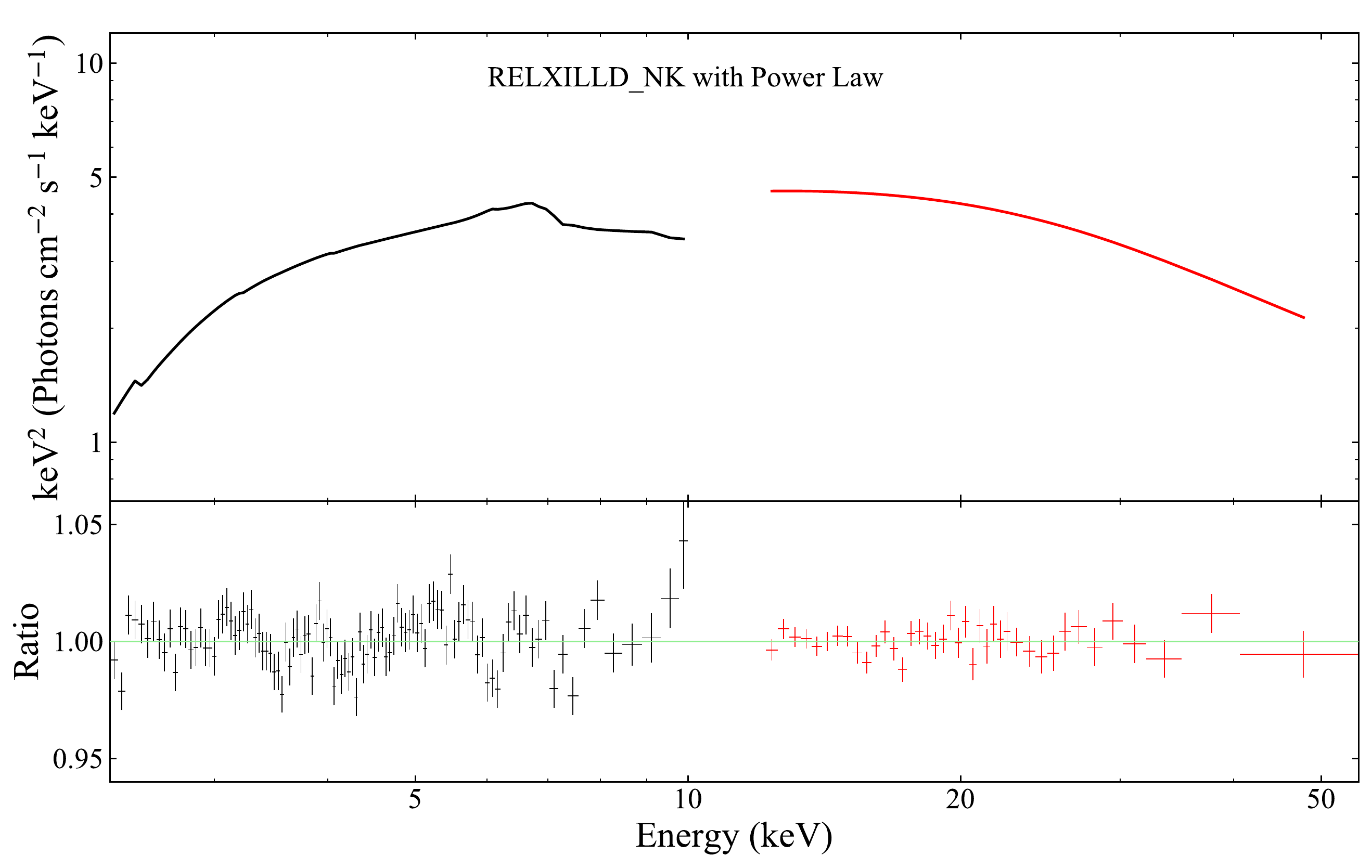}
\hspace{0.4cm}
\includegraphics[width=0.45\textwidth]{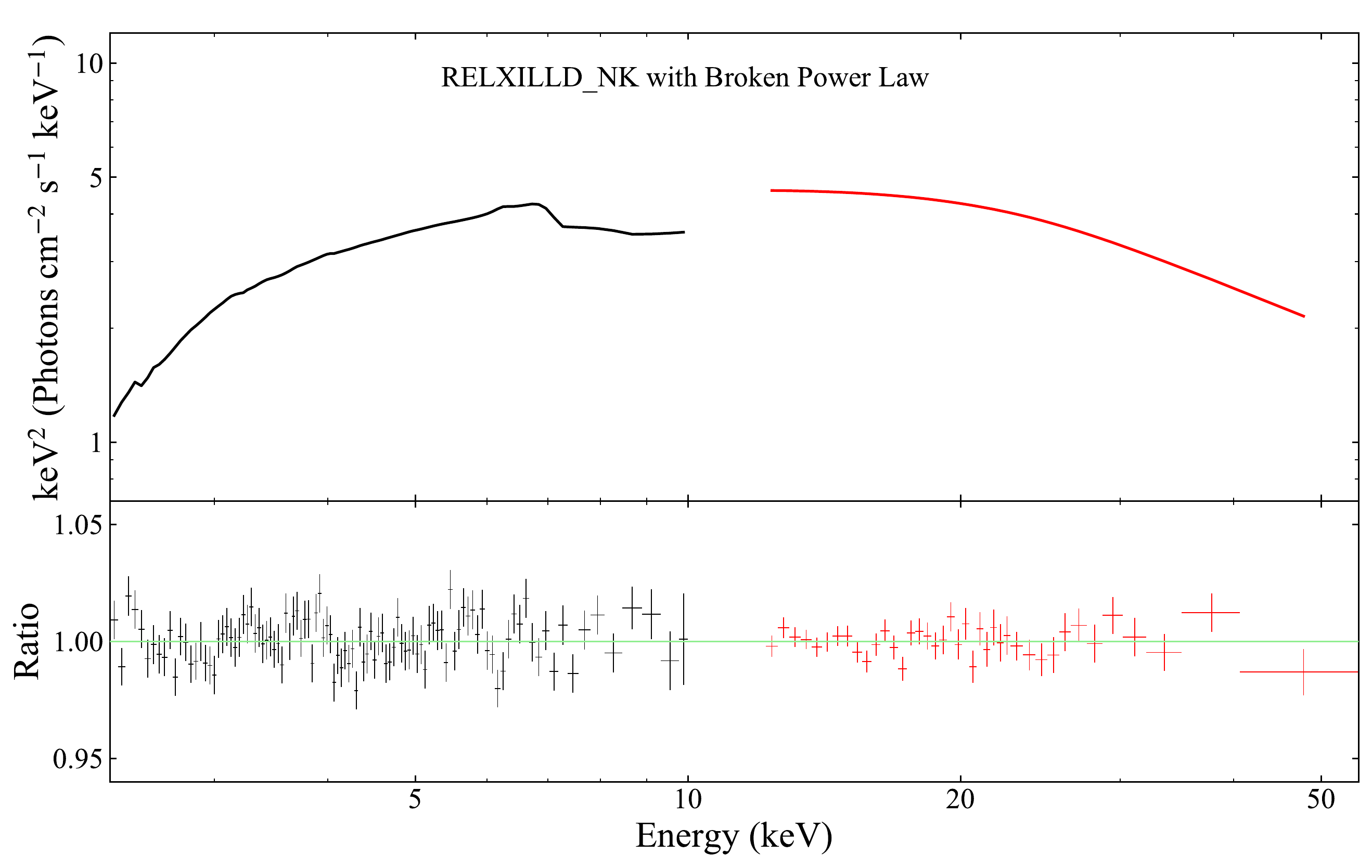}
\end{center}
%\vspace{-1.0cm}
\caption{Fit results for the Johannsen model with $\alpha_{13}$ free for {\sc relxill\_nk},  {\sc relxillCp\_nk}, and {\sc relxillD\_nk}. For every flavor, we have two models for the emissivity profile: simple power-law (left column) and broken power-law (right column). In every panel, the top quadrant shows the best-fit model and the bottom quadrant shows the data to best-fit model ration. XIS1 data/fit in black and HXD/PIN data/fit in red. \label{f-ratio}}
\vspace{0.4cm}
\end{figure*}

\begin{figure*}[t]
\begin{center}
\vspace{0.3cm}
\includegraphics[width=0.49\textwidth]{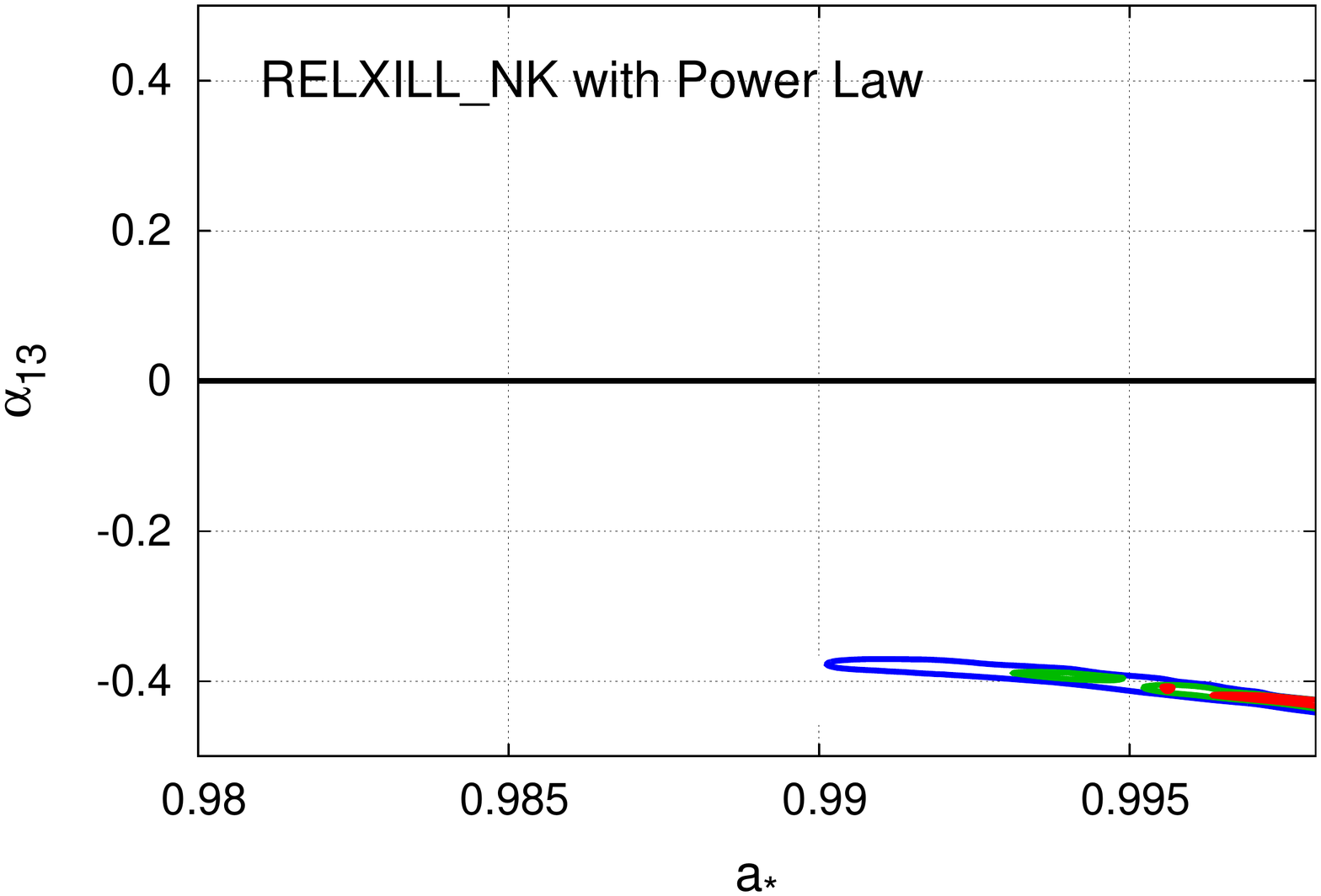}
%\hspace{0.2cm}
\includegraphics[width=0.49\textwidth]{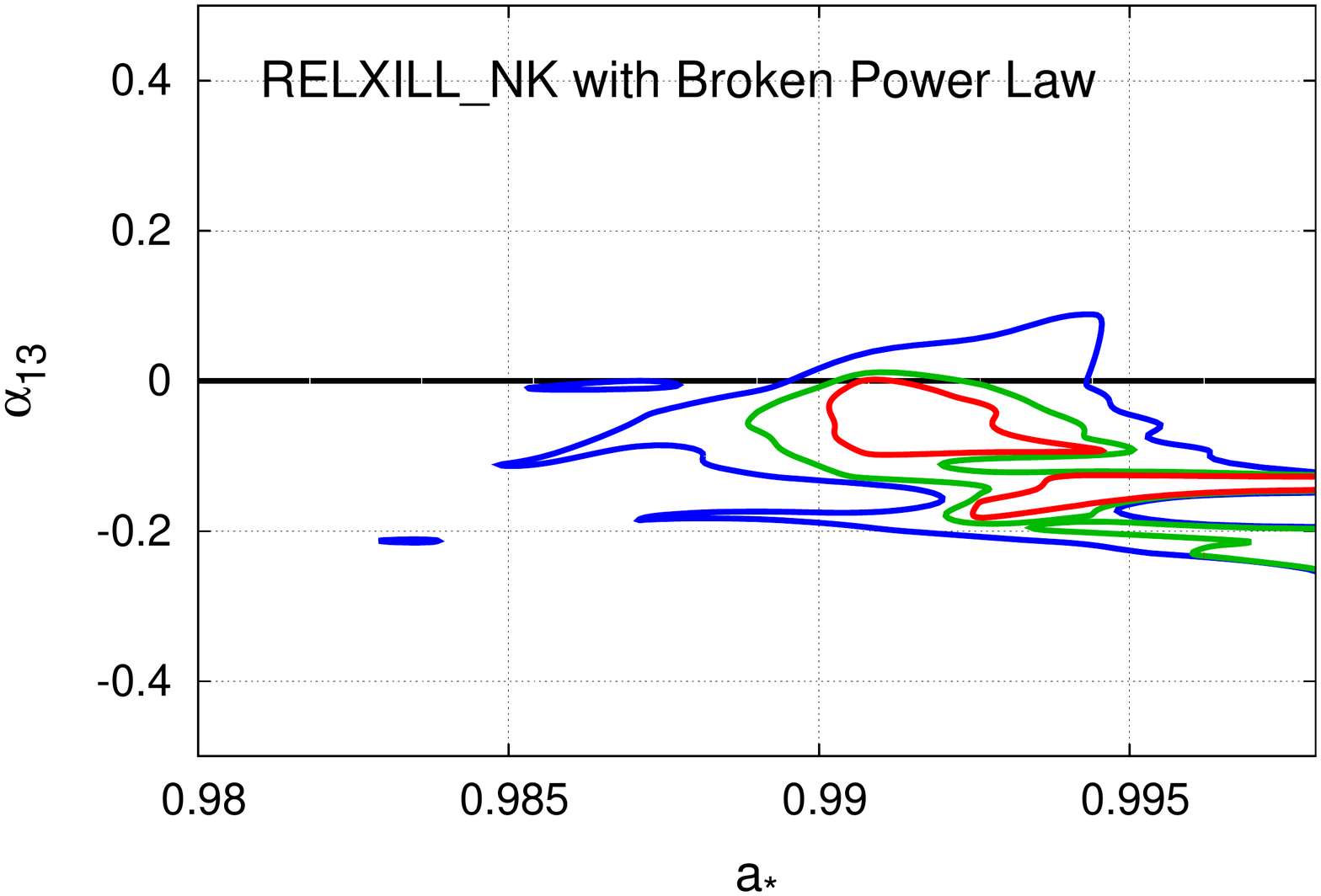} 
\vspace{-1.1cm} \\
\includegraphics[width=0.49\textwidth]{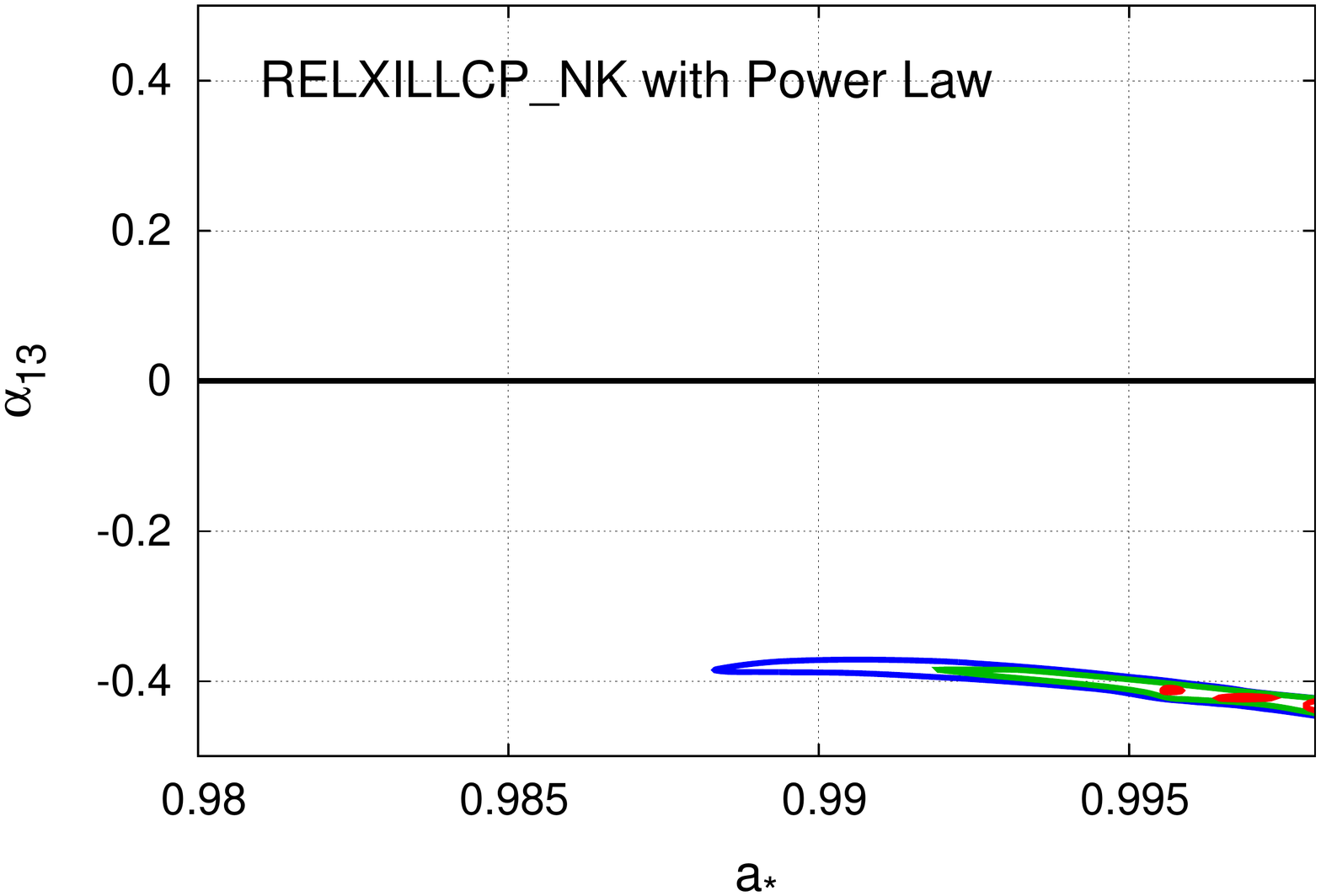}
%\hspace{0.2cm}
\includegraphics[width=0.49\textwidth]{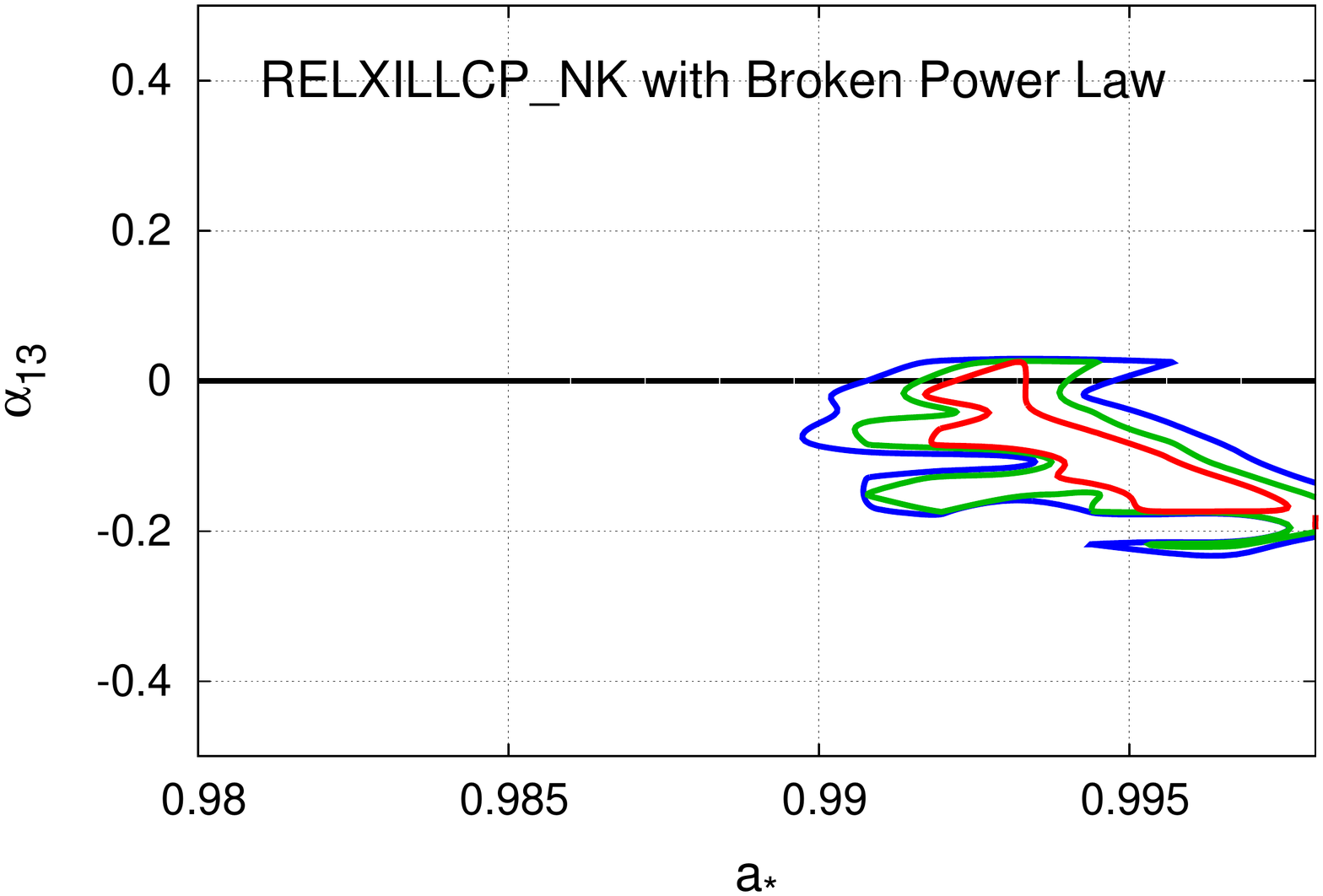} 
\vspace{-1.1cm} \\
\includegraphics[width=0.49\textwidth]{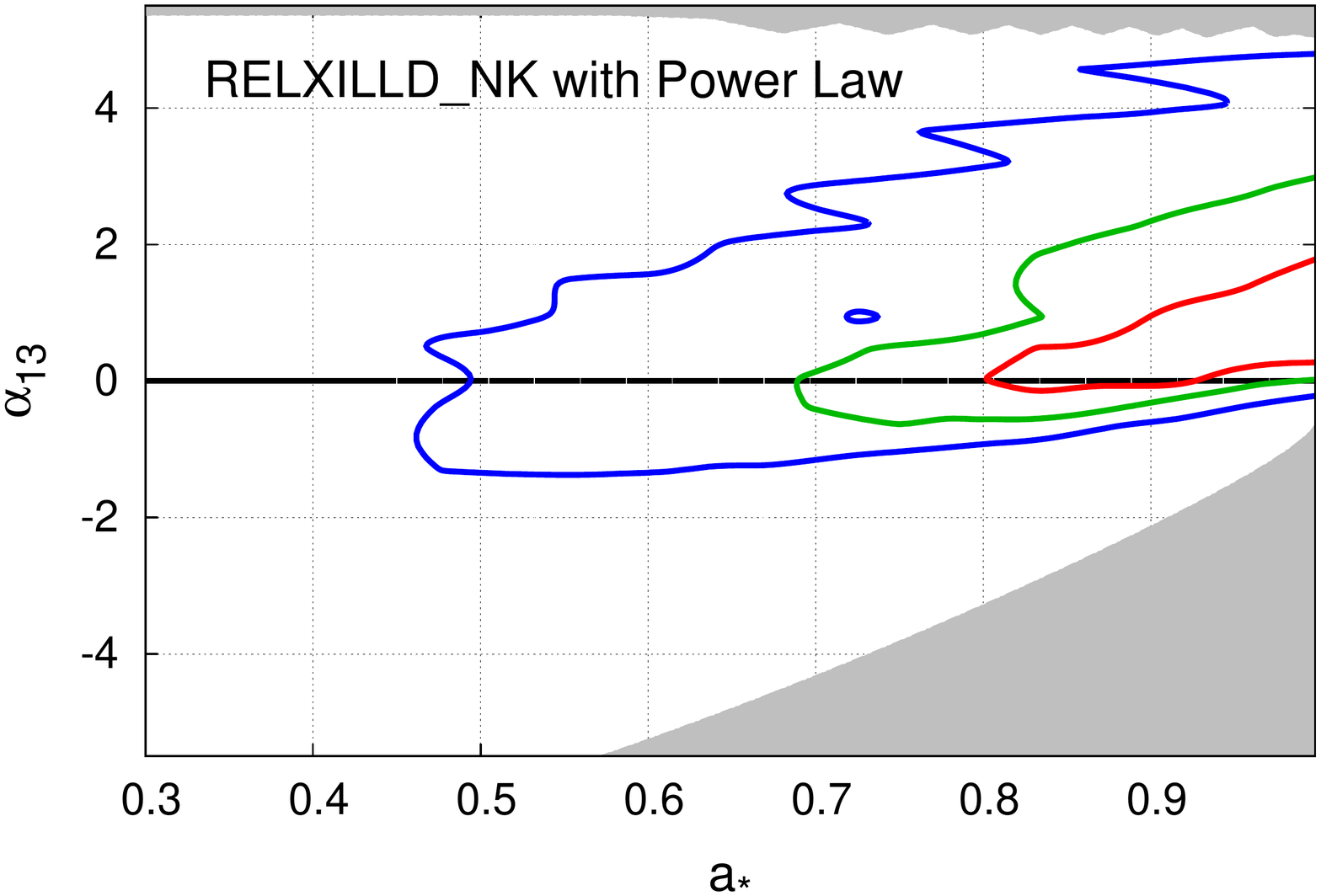}
%\hspace{0.2cm}
\includegraphics[width=0.49\textwidth]{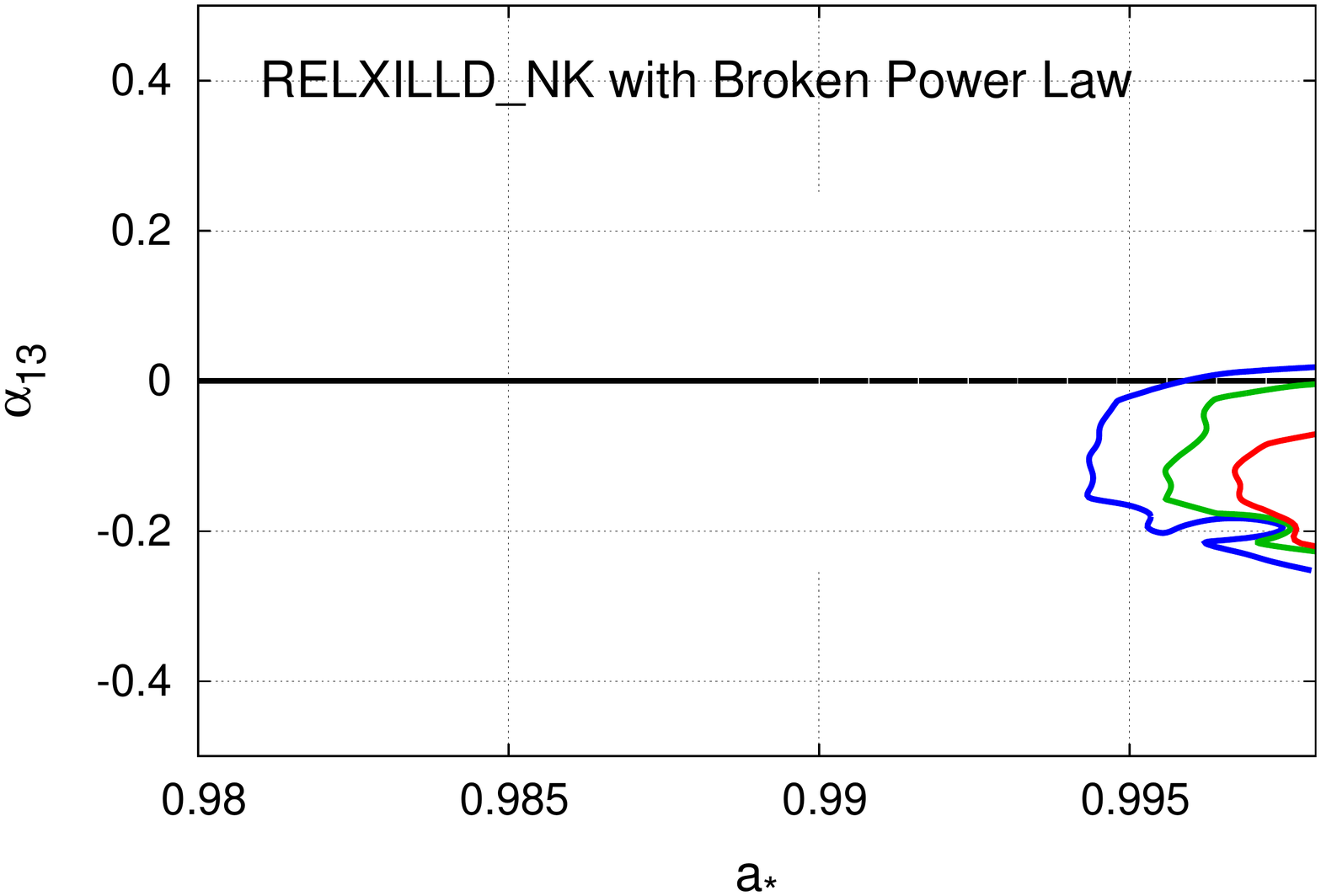}
\end{center}
\vspace{-1.0cm}
\caption{Constraints on the spin parameter $a_*$ and the deformation parameter $\alpha_{13}$ employing different {\sc relxill} flavors: the default model {\sc relxill\_nk} (top panels), {\sc relxillCp\_nk} (central panels), and {\sc relxillD\_nk} (bottom panels). In the left panels, the intensity profile of the reflection spectrum is modeled with a simple power-law, in the right panels we use a broken power-law with both inner and outer emissivity indices free in the fit. The red, green, and blue curves correspond, respectively, to the 68\%, 90\%, and 99\% confidence contours for two relevant parameters. Note that these constraints are obtained by marginalizing over all other free parameters of the fit. The gray region is not analyzed in our study because the spacetime is not regular there [see Eq.~(\ref{eq-constraints}) in the appendix]. \label{f-a13}}
\vspace{0.4cm}
\end{figure*}

\begin{figure*}[t]
\begin{center}
\vspace{0.3cm}
\includegraphics[width=0.49\textwidth]{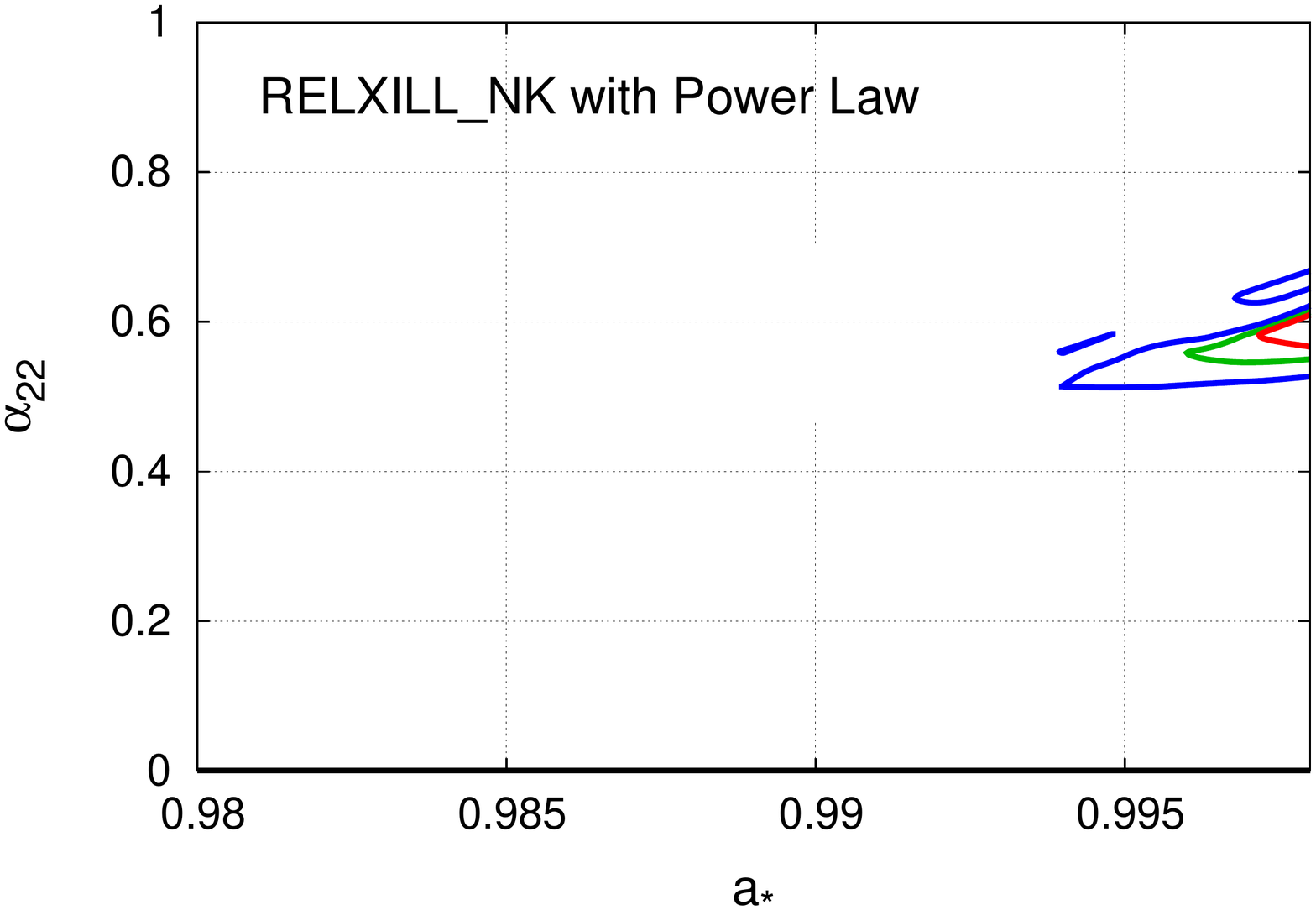}
%\hspace{0.2cm}
\includegraphics[width=0.49\textwidth]{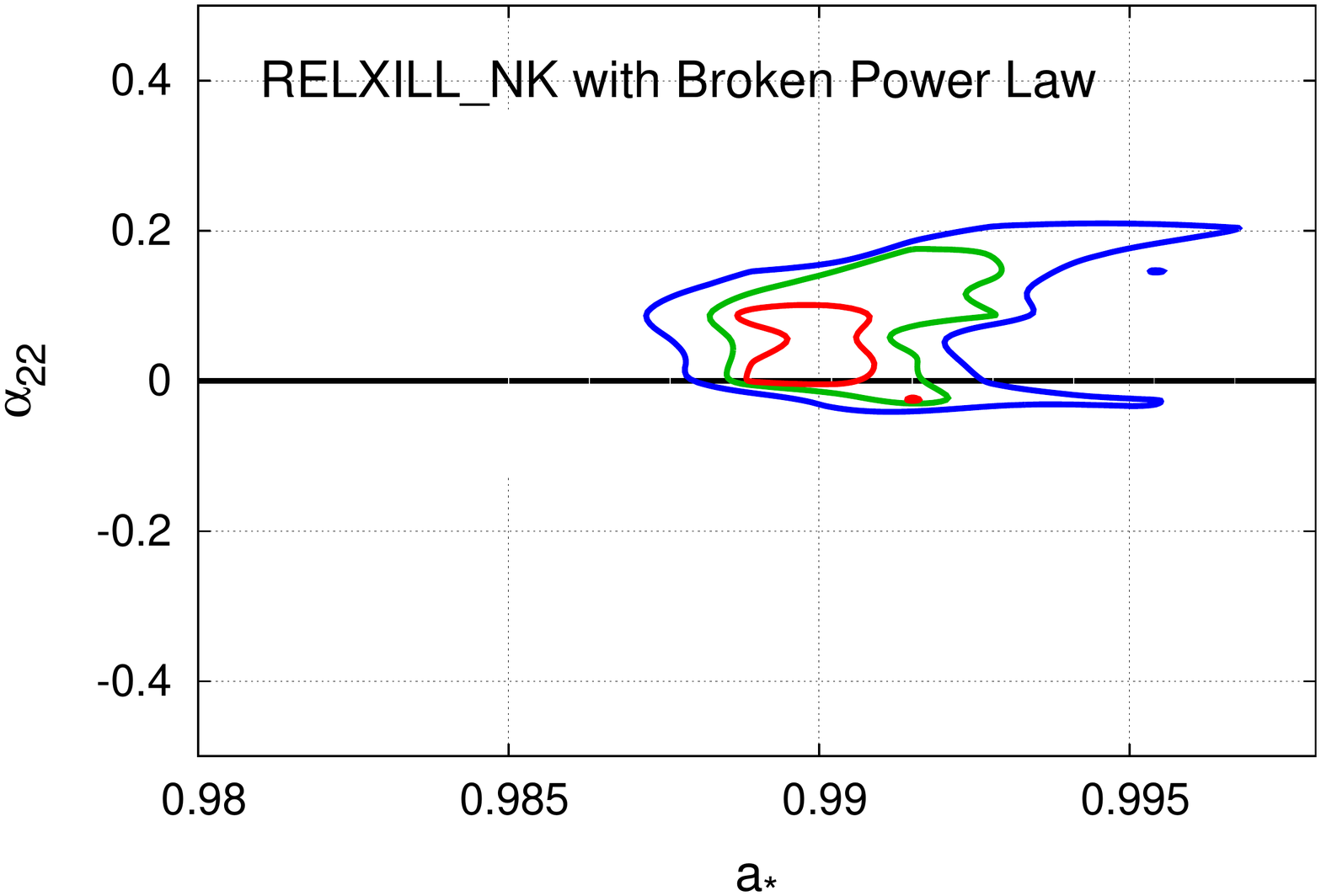} 
\vspace{-1.1cm} \\
\includegraphics[width=0.49\textwidth]{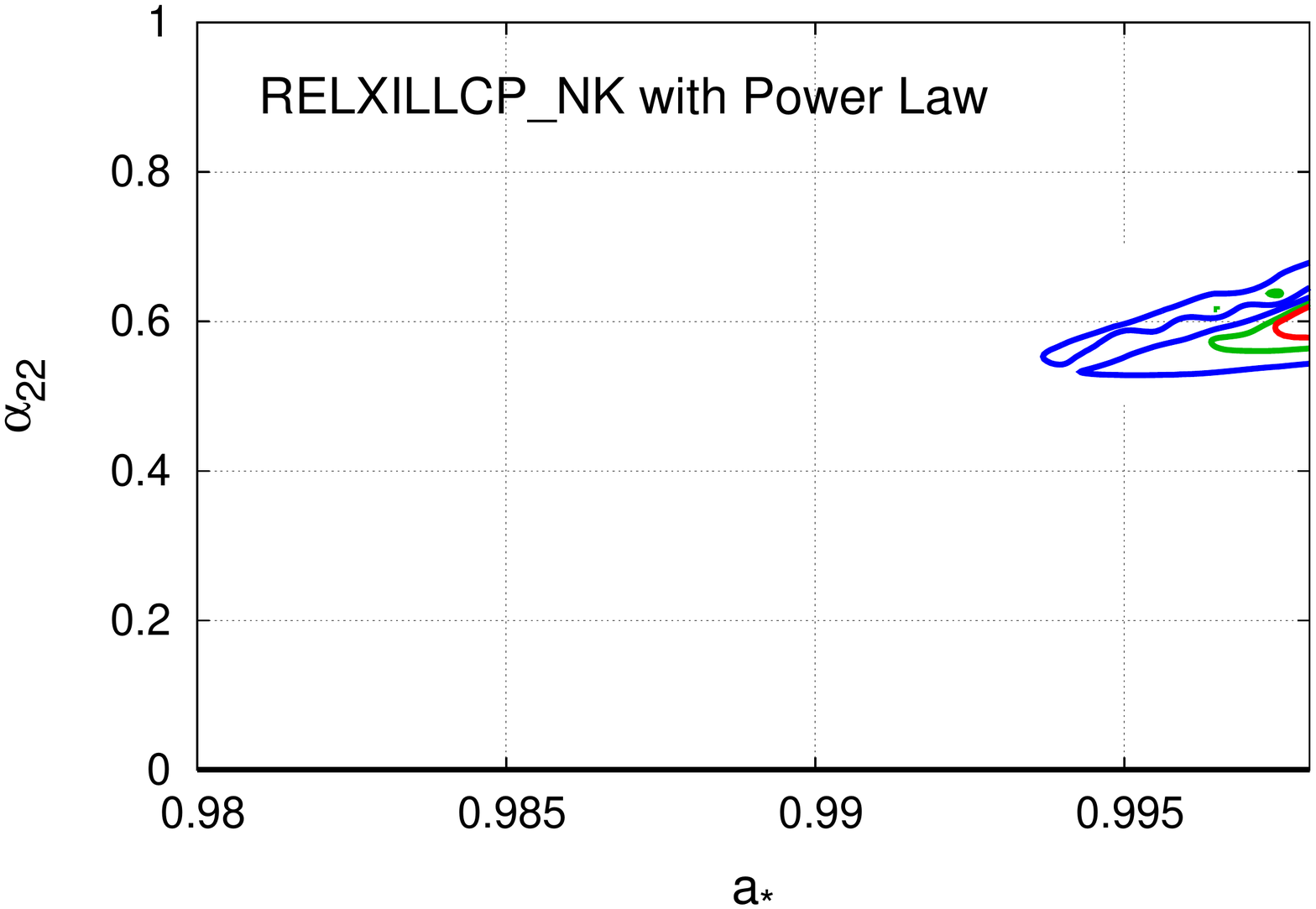}
%\hspace{0.2cm}
\includegraphics[width=0.49\textwidth]{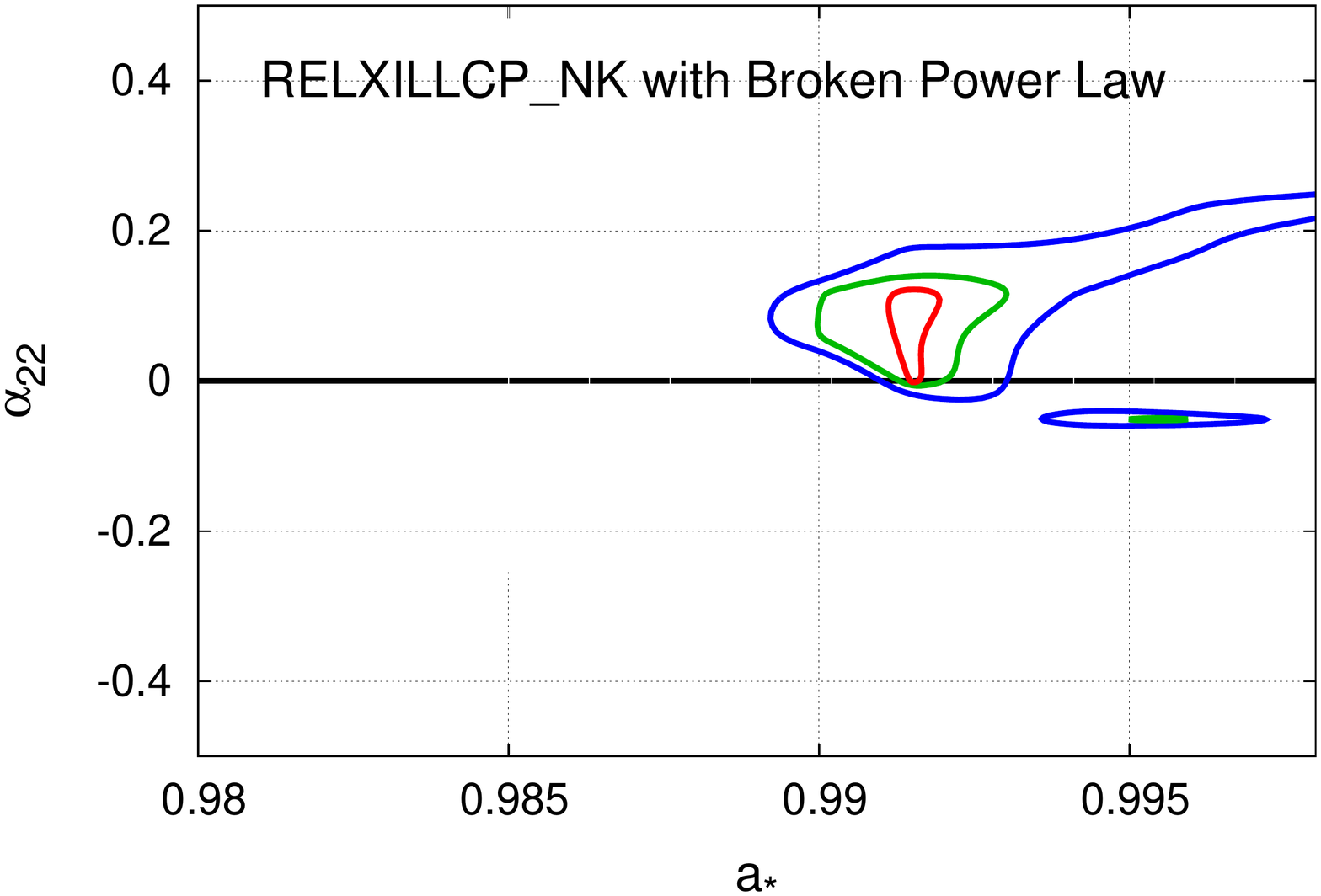} 
\vspace{-1.1cm} \\
\includegraphics[width=0.49\textwidth]{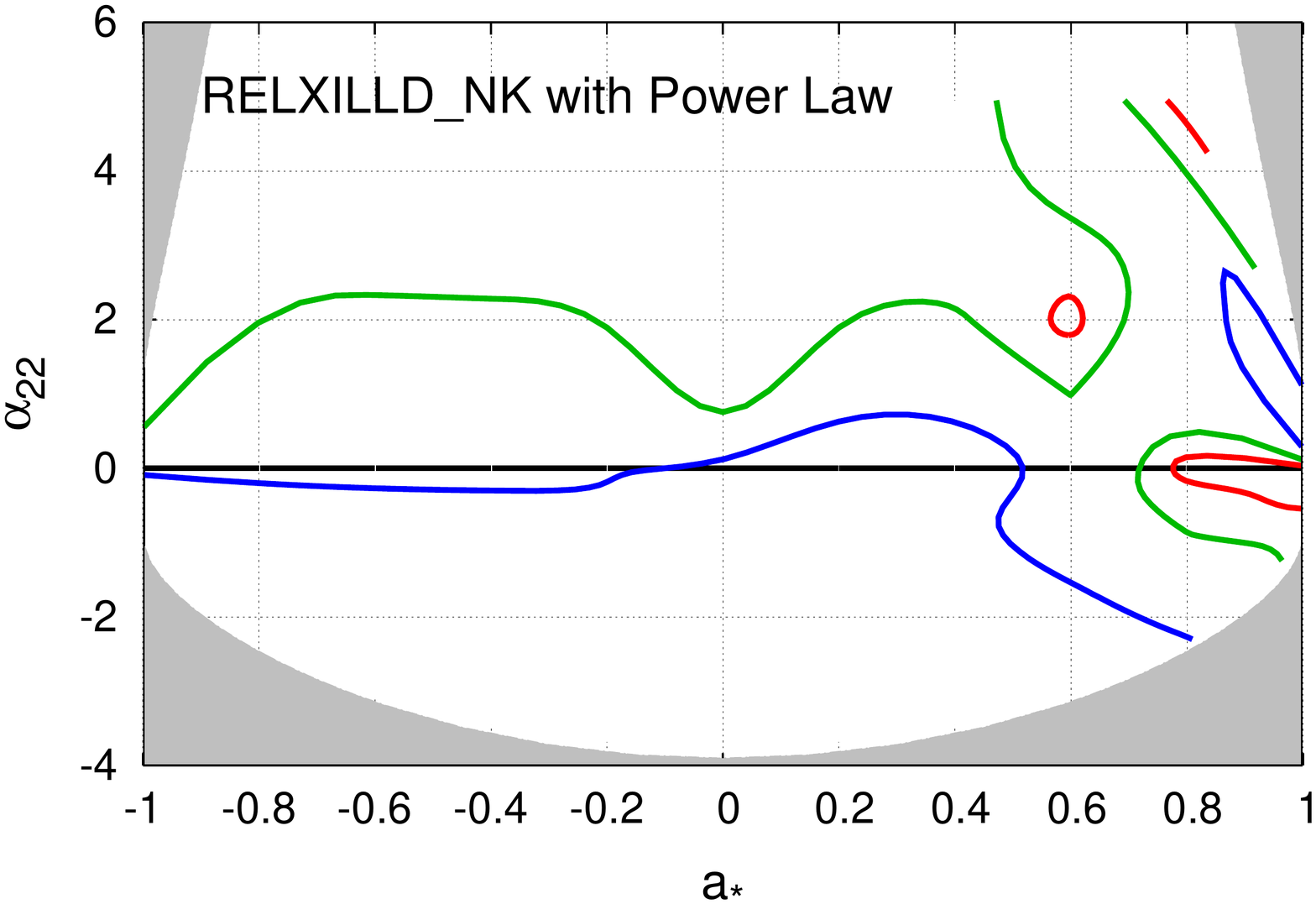}
%\hspace{0.2cm}
\includegraphics[width=0.49\textwidth]{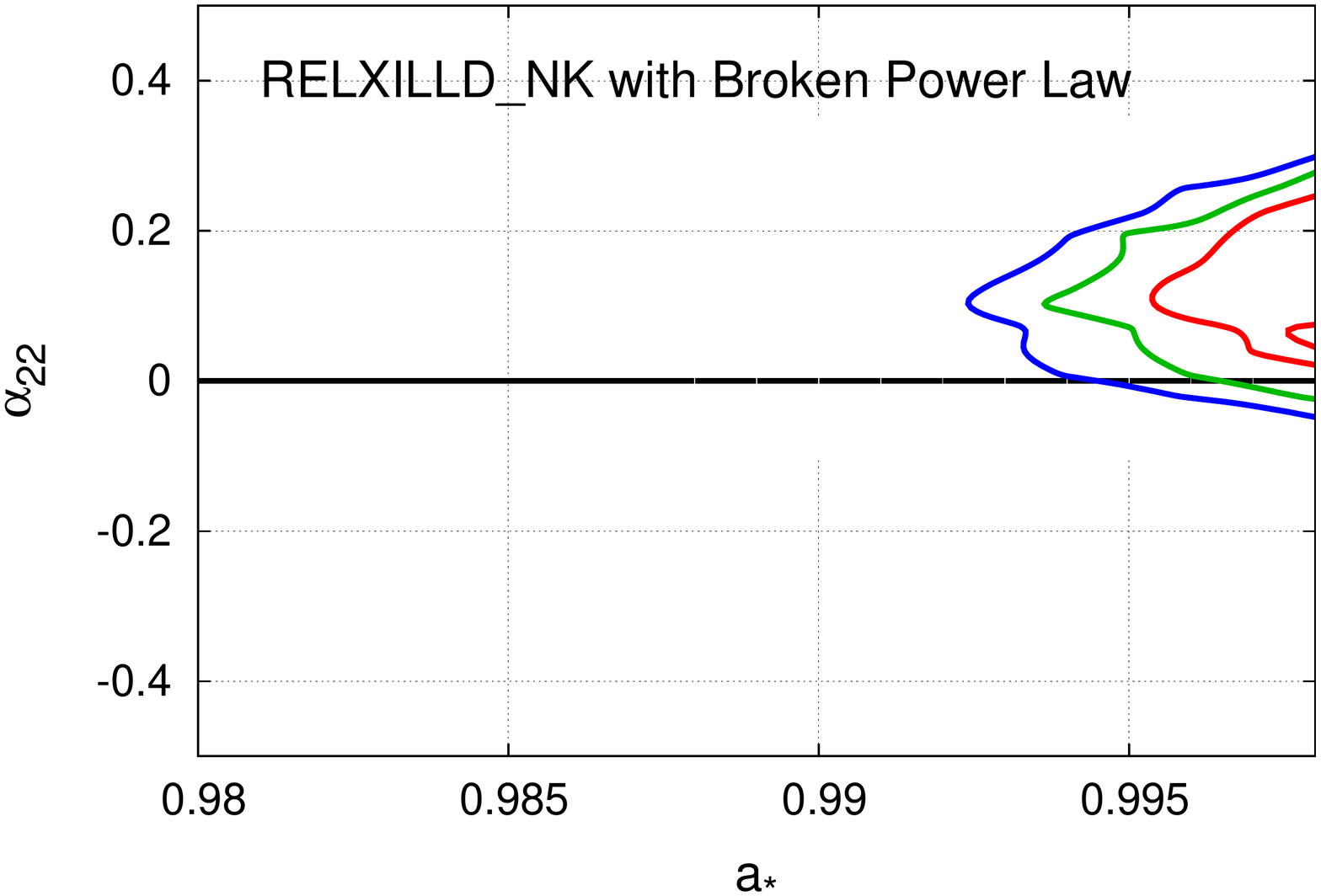}
\end{center}
\vspace{-1.0cm}
\caption{As in Fig.~\ref{f-a13} with the deformation parameter $\alpha_{22}$ allowed to vary and all other deformation parameters set to zero. \label{f-a22}}
\vspace{0.4cm}
\end{figure*}

\begin{figure*}[t]
\begin{center}
\vspace{0.3cm}
\includegraphics[width=0.49\textwidth]{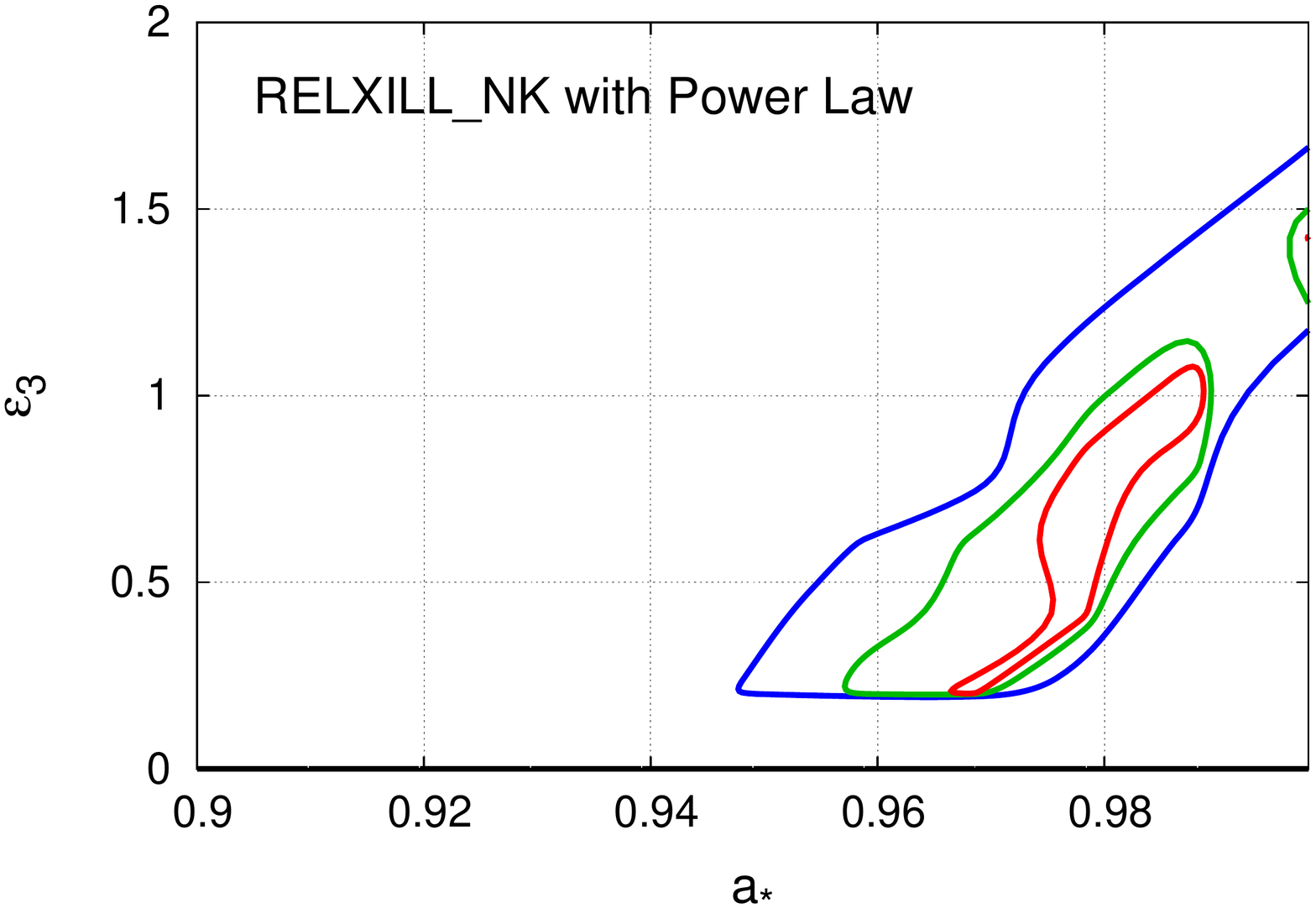}
%\hspace{0.2cm}
\includegraphics[width=0.49\textwidth]{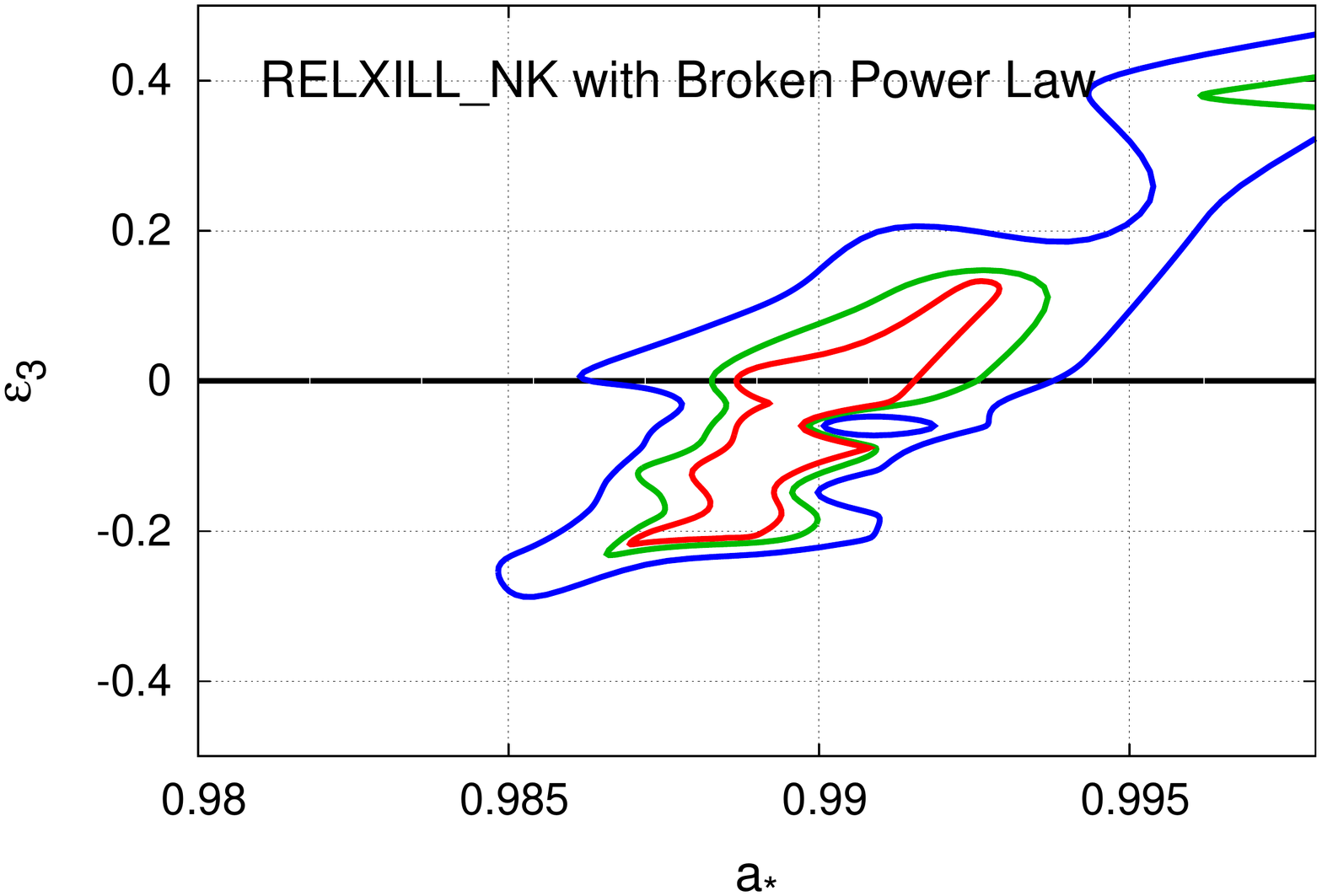} 
\vspace{-1.1cm} \\
\includegraphics[width=0.49\textwidth]{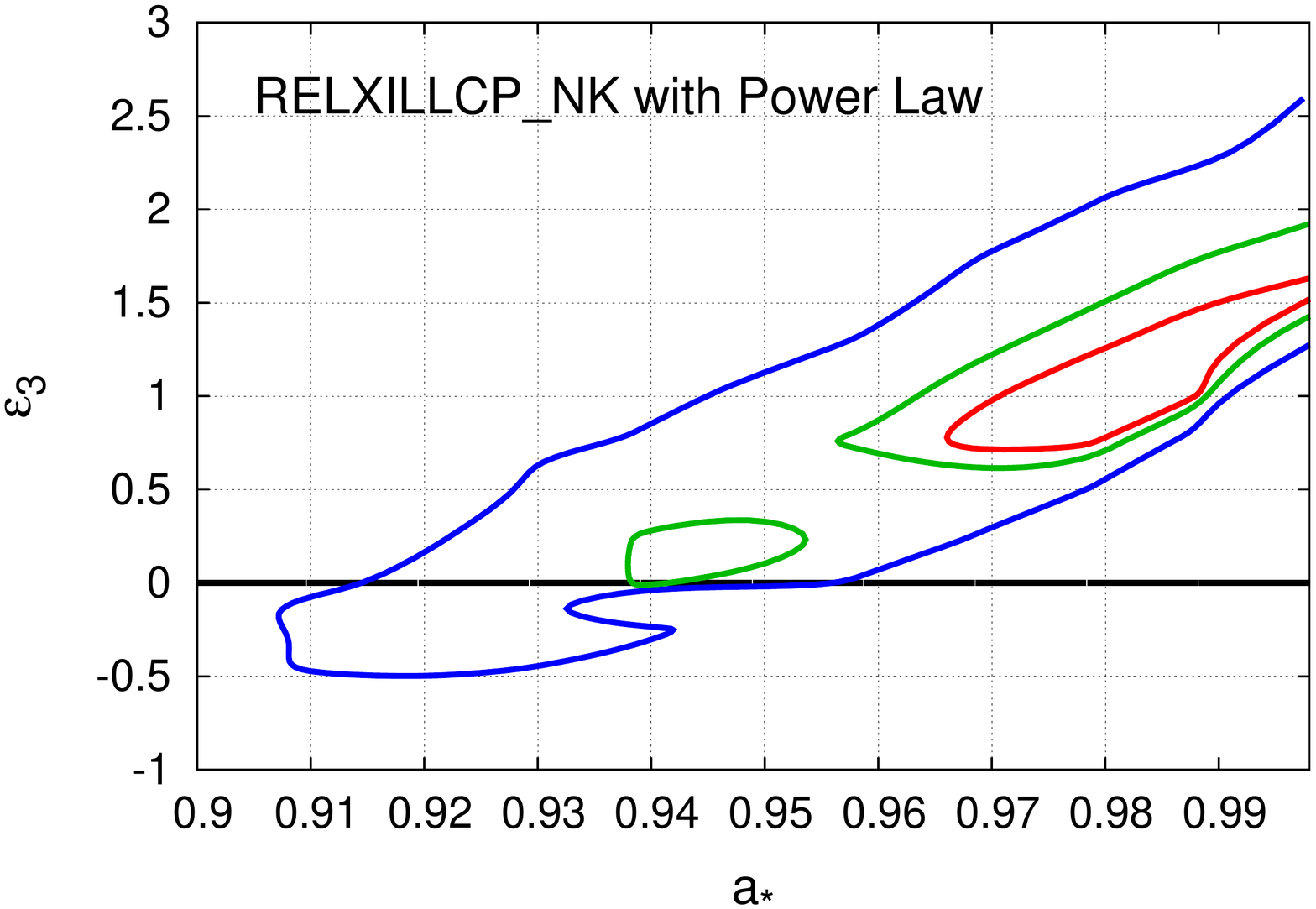}
%\hspace{0.2cm}
\includegraphics[width=0.49\textwidth]{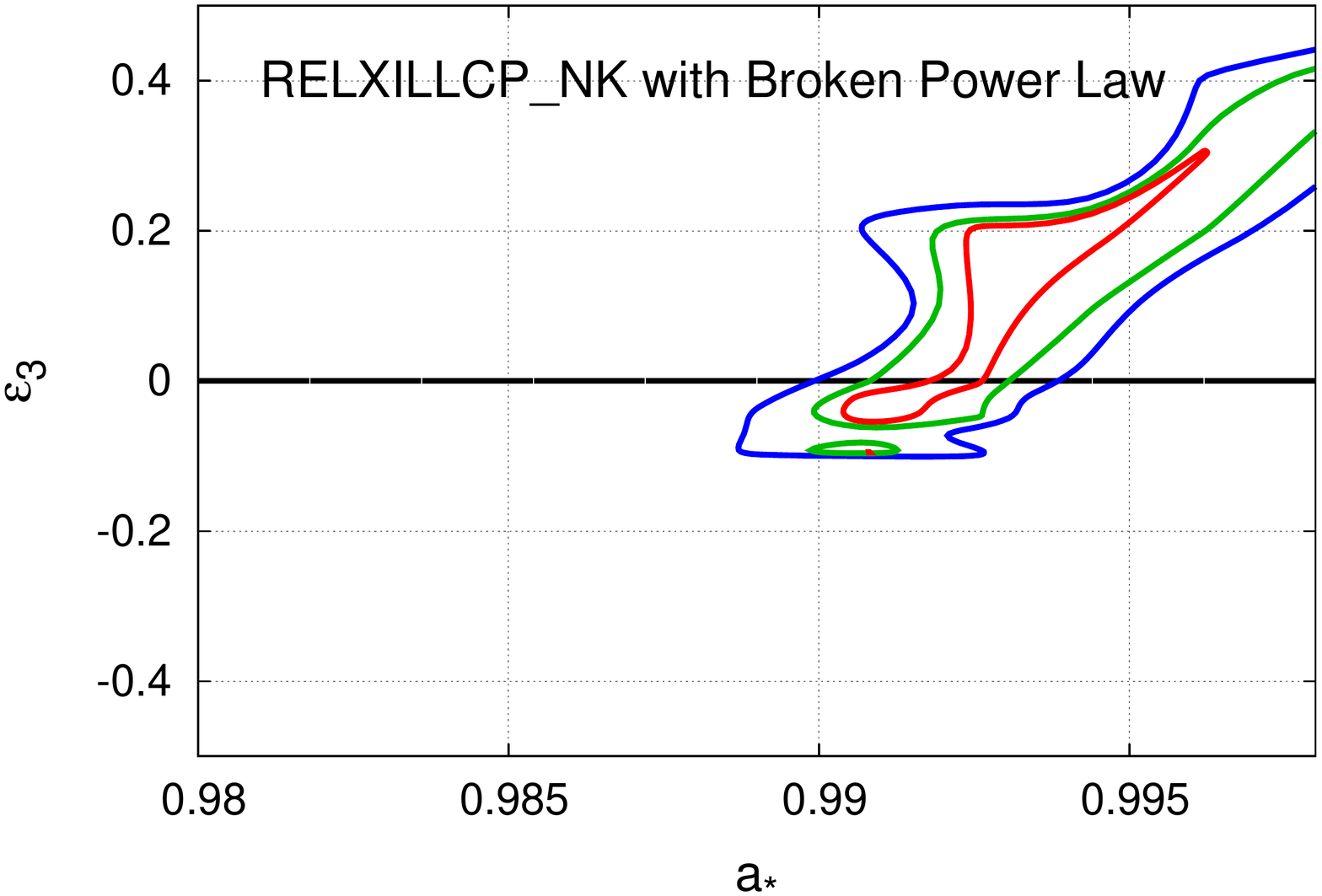} 
\vspace{-1.1cm} \\
\includegraphics[width=0.49\textwidth]{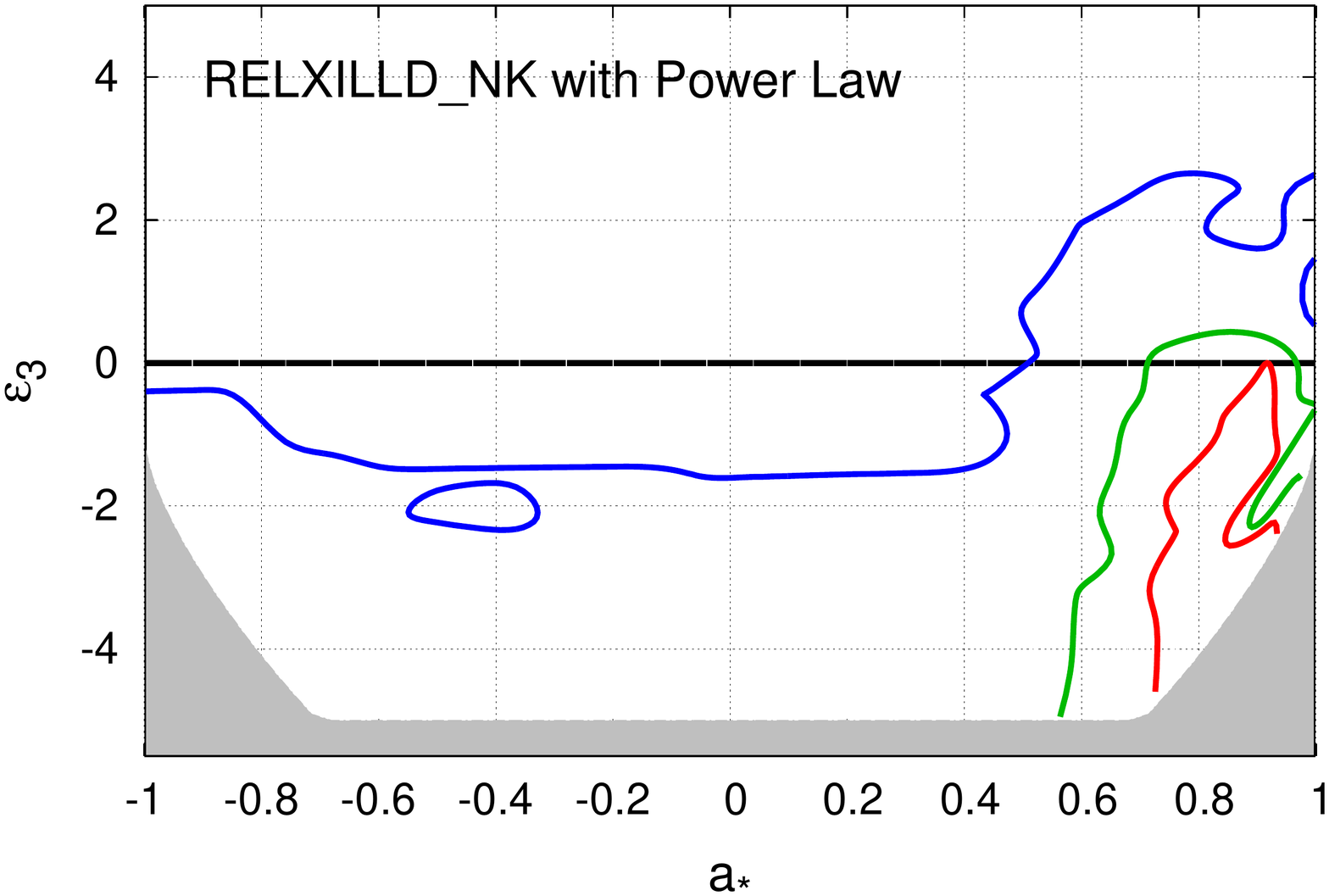}
%\hspace{0.2cm}
\includegraphics[width=0.49\textwidth]{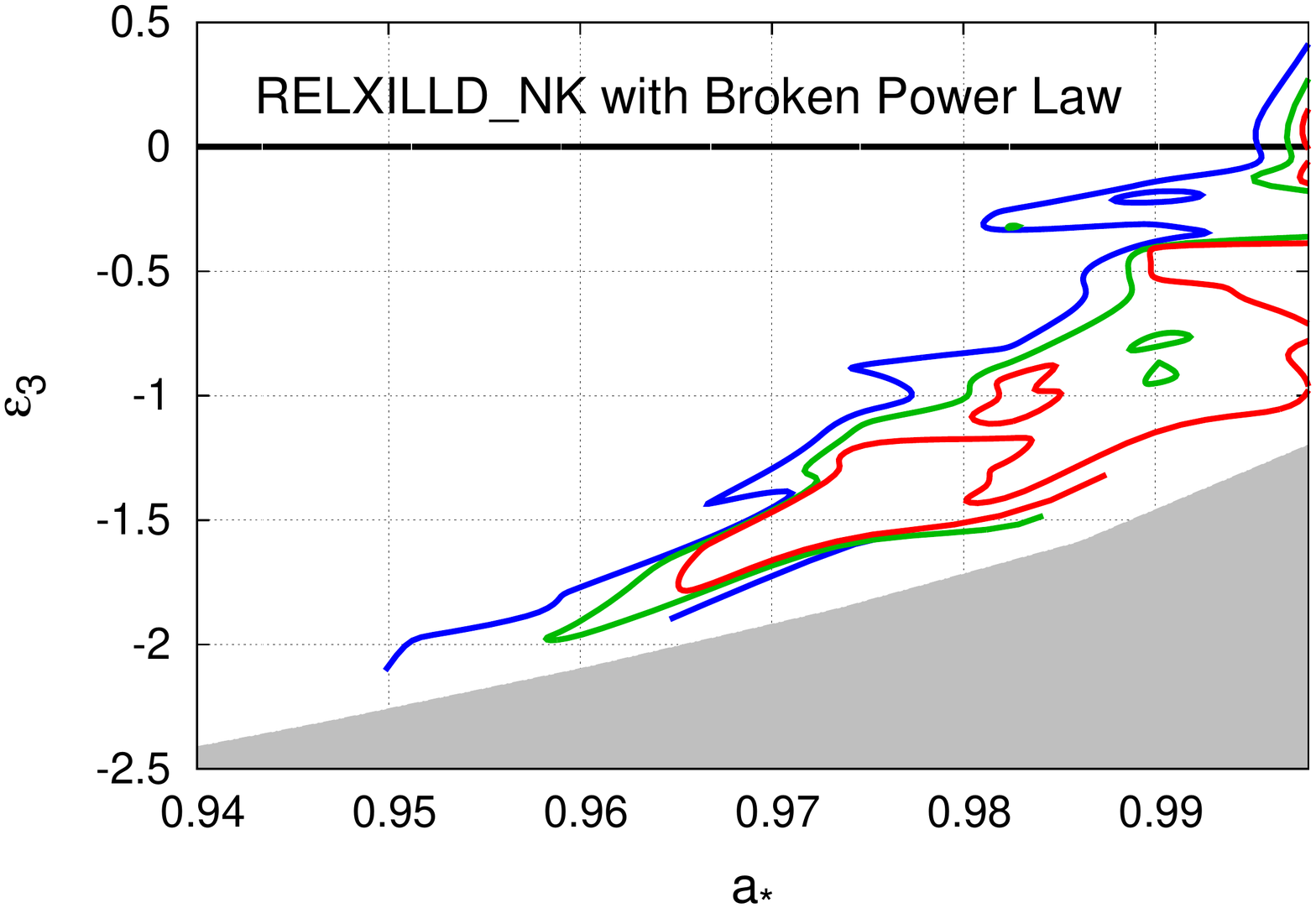}
\end{center}
\vspace{-1.0cm}
\caption{As in Fig.~\ref{f-a13} with the deformation parameter $\epsilon_3$ allowed to vary and all other deformation parameters set to zero. \label{f-e3}}
\vspace{0.4cm}
\end{figure*}

\section{Spectral analysis}\label{s-ana}

The spectrum of the GRS~1915+105 observed by \textsl{Suzaku} in 2007 seems to be quite simple and already well modeled with absorbed coronal and reflection components. This is also the conclusion in \citet{blum}. In particular, we note that we do not need the disk's thermal component. In the Xspec language, the model is {\sc tbabs$\times$relxill\_nk}. {\sc tbabs} describes the Galactic absorption~\citep{wilms}, and the hydrogen column density $N_{\rm H}$ is left free in all the fits. {\sc relxill\_nk} is our relativistic reflection model and includes the coronal spectrum~\citep{noim1,noim2}. The model employs the Johannsen metric~\citep{tj}, which is briefly reviewed in Appendix~\ref{a-metric}. With respect to the standard {\sc relxill}~\citep{rx1,rx2,rx3}, we have three extra parameters, called, respectively, $\alpha_{13}$, $\alpha_{22}$, and $\epsilon_3$, which quantify possible deviations from the Kerr solution. The Kerr metric is recovered when these deformation parameters vanish. Measuring the value of these deformation parameters thus tests the Kerr hypothesis. Note that the current version of {\sc relxill\_nk} only allows for the measurement of one of these deformation parameters assuming the other two vanish. Here, for {\sc relxill\_nk}, we do not only assume the default model {\sc relxill\_nk}, but we also include the flavors {\sc relxillCp\_nk} and {\sc relxillD\_nk}.

We have four ``main'' models: Kerr spacetime ($\alpha_{13} = \alpha_{22} = \epsilon_3 = 0$), Johannsen spacetime with free $\alpha_{13}$, Johannsen spacetime with free $\alpha_{22}$, and Johannsen spacetime with free $\epsilon_3$. For each of these four cases, we fit the data with {\sc relxill\_nk}, {\sc relxillCp\_nk}, and {\sc relxillD\_nk}. For every flavor, we assume two possible emissivity profiles for the reflection spectrum: simple power-law and broken power-law with both inner and outer emissivity indices and breaking radius free. The results of our fits are shown in Tabs.~\ref{t-kerr}, \ref{t-a13}, \ref{t-a22}, and \ref{t-e3}, respectively for the model with the Kerr spacetime, $\alpha_{13}$, $\alpha_{22}$, and $\epsilon_3$. The best-fit models and the data to best-fit model ratios for the Johannsen spacetime with free $\alpha_{13}$ are shown in Fig.~\ref{f-ratio} (the cases for the Kerr spacetime and Johannsen spacetime with free $\alpha_{22}$ or $\epsilon_3$ are similar). The constraints on the spin vs deformation parameter plane are of particular interest for us. These are shown in Fig.~\ref{f-a13} ($a_*$ vs $\alpha_{13}$), Fig.~\ref{f-a22} ($a_*$ vs $\alpha_{22}$), and Fig.~\ref{f-e3} ($a_*$ vs $\epsilon_{3}$). The red, green, and blue curves correspond, respectively, to the 68\%, 90\%, and 99\% confidence contours for two relevant parameters.

%%%%%%%%%%%%%%%%%%%%%%%%%%%%%%%

\section{Discussion and conclusions}\label{s-dis}

Let us start by discussing the Kerr models. The best-fit values are shown in Tab.~\ref{t-kerr}. Unfortunately, we cannot easily compare our results with those in \citet{blum} because \citet{blum} use a different reflection model and employs different assumptions, freezing some model parameters that here we decide to have free. Generally speaking, we find that the stellar-mass black hole must rotate fast, in agreement with the other measurements in the literature~\citep{grs1,blum,grs2}. Note, however, that when we model the emissivity profile with a broken power-law, the spin is very high ($a_* > 0.99$) and this is true for every flavor. Assuming a simple power-law, we find a moderately high spin. Moreover, in the case of a simple power-law, we find a high value of the emissivity index (with the exception of {\sc relxillD\_nk}). When we employ a broken power-law, we always find a very high $q_{\rm in}$ and a very low $q_{\rm out}$. The fit with {\sc relxillD\_nk} and a simple power-law is quite suspicious: we find quite a low emissivity index, the inclination of the disk is stuck at the maximum (which was set to 80$^\circ$ in this analysis, because our model becomes inaccurate for higher values), and the ionization of the disk is clearly too low. In {\sc relxillD\_nk}, the parameter associated to the disk electron density, $\log{\rm N}$, can only vary from 15 to 19, so high disk densities are not included; our best-fit value is somewhat lower than what is expected theoretically~\citep{SZ94,javier2} and found in GX~339--4 in \citet{jiachen2}.

The results for the Johannsen spacetime with free $\alpha_{13}$ are shown in Tab.~\ref{t-a13}. Fig.~\ref{f-a13} shows the constraints on the spin vs. deformation parameter plane. Ignoring the case with {\sc relxillD\_nk} and a simple power-law, which should be discussed separately, the other fits show a clear trend. With a simple power-law, we do not recover the Kerr solution with a very high confidence level. When we employ a broken power-law, we get a better fit ($\Delta\chi^2 \sim 30$-50) and we recover the Kerr solution. We are tempted to conclude that: $i)$ the choice of the model of the intensity profile is important, $ii)$ a simple power-law is not adequate to model the reflection spectrum of this \textsl{Suzaku} observation, and $iii)$ X-ray reflection spectroscopy can potentially measure both the intensity profile and the spacetime metric. Note also that the constraints on $\alpha_{13}$ obtained with the broken power-law are quite competitive when compared with previous results from other sources~\citep{noi4,noi5,noi6}. This was definitively not the situation found in \citet{yuexin} with the \textsl{NuSTAR} data of GRS~1915+105.

Concerning the model with {\sc relxillD\_nk} and a simple power-law, we note that, like in the case of the Kerr model, the best-fit values are quite suspicious. The emissivity index and the ionization parameter are unreasonably low and the inclination angle of the disk is unreasonably high. Actually we recover the Kerr metric at a low confidence level, but just because the uncertainty is large.

For the models with the Johannsen metric and free $\alpha_{22}$ or $\epsilon_3$ we find similar results. The fit with {\sc relxillD\_nk} and a simple power-law always provides quite unphysical values of some model parameters, so we can argue it cannot be the right model. For the other fits, when we employ a simple power-law, we do not recover the Kerr solution (or we marginally recover the Kerr solution, see the case with $\epsilon_3$ and {\sc relxillCp\_nk}). When we employ a broken power-law, we confirm the Kerr hypothesis with quite competitive constraints.

We argue that there are two important differences between the study reported here and that in \citet{yuexin}: $i)$ \textsl{Suzaku} has a much better energy resolution near the iron line than \textsl{NuSTAR} and tests of the Kerr metric probably require high energy resolution near the iron line, and $ii)$ here we do not see any thermal component from the disk, while a thermal component was necessary in the \textsl{NuSTAR} data, suggesting that the disk temperature here is lower. Note that in our model the reflection spectrum at the emission point is that of {\sc xillver}, where the calculations are done assuming a cold disk and therefore it is perfectly understandable that the model is more suitable for disks with lower temperature. Note also that our discussion could include the study of Cygnus~X-1 in the soft state with \textsl{NuSTAR} data reported in \citet{honghui}. Like in \citet{yuexin}, even in \citet{honghui} we were not able to test the Kerr hypothesis and the uncertainties were large, supporting our idea that our tests need cold disks and high energy resolution near the iron line.

Last, we note that the modeling effects studied in the present paper, i.e. disk density and corona spectrum, are not the only ones that can have an impact on the estimate of the deformation parameters. Our reflection model has a number of simplifications that inevitably can introduce systematic errors in the final measurements; see \citet{honghui} for the list of simplifications in the current version of the model. For instance, our model assumes an infinitesimally thin disk, while real disks have a finite thickness~\citep{2018ApJ...855..120T}. The ionization parameter $\xi$ is assumed to be constant over the disk, while it is expected to vary over radius according to the X-ray flux from the corona and the disk density~\citep{2012A&A...545A.106S,2019MNRAS.485..239K,2019MNRAS.488..324I}. Here the emissivity profile is modeled with a power-law or a broken power-law, which is clearly quite a crude approximation and does not permit one to calculate all the relativistic effects; the emissivity profile can be self-consistently calculated from a specific coronal geometry, like in~\citet{rx1}. As the quality of the available X-ray data improve, it will be mandatory to improve our theoretical model and study the impact of model simplifications in order to try to perform precision tests of the Kerr metric with X-ray reflection spectroscopy.

\vspace{0.3cm}

%%%%%%%%%%%%%%%%%%%%%%%%%%%%%%%

{\bf Acknowledgments --}
This work was supported by the Innovation Program of the Shanghai Municipal Education Commission, Grant No.~2019-01-07-00-07-E00035, and Fudan University, Grant No.~IDH1512060. Y.Z. also acknowledges support from the Fudan Undergraduate Research Opportunities Program (FDUROP). A.B.A. also acknowledges support from the Shanghai Government Scholarship (SGS). S.N. acknowledges support from the Alexander von Humboldt Foundation and the Excellence Initiative at Eberhard-Karls Universit\"at T\"ubingen.

%%%%%%%%%%%%%%%%%%%%%%%%%%%%%%%

\appendix

\section{The {\sc relxill\_nk} model}\label{a-metric}

{\sc relxill}~\citep{rx1,rx2,rx3} is currently the most advanced relativistic reflection model in the Kerr metric. {\sc relxill\_nk}~\citep{noim1,noim2} is an extension of {\sc relxill}: the model does not assume the Kerr background and employs a parametric black hole metric. The latter is not a black hole solution of some particular theory of gravity but a black hole metric obtained by deforming the Kerr solution. The metric is specified by the mass $M$ and the spin angular momentum $J$ of the black hole as well as by a number of deformation parameters, which are introduced to describe deviations from the Kerr geometry. The Kerr metric is exactly recovered when all deformation parameters vanish. The strategy is thus to fit the data with this metric, estimate the value of the deformation parameters with some statistical tool, and check whether the data require vanishing deformation parameters in order to satisfy the Kerr hypothesis.

In the analysis presented in this paper, we have considered the Johannsen metric~\citep{tj}. In Boyer-Lindquist-like coordinates, the line element of the Johannsen metric reads
\be\label{eq-jm}
ds^2 &=&-\frac{\tilde{\Sigma}\left(\Delta-a^2A_2^2\sin^2\theta\right)}{B^2}dt^2
+\frac{\tilde{\Sigma}}{\Delta}dr^2+\tilde{\Sigma} d\theta^2 
-\frac{2a\left[\left(r^2+a^2\right)A_1A_2-\Delta\right]\tilde{\Sigma}\sin^2\theta}{B^2}dtd\phi \nonumber\\
&&+\frac{\left[\left(r^2+a^2\right)^2A_1^2-a^2\Delta\sin^2\theta\right]\tilde{\Sigma}\sin^2\theta}{B^2}d\phi^2
\ee
where $M$ is the black hole mass, $a = J/M$, $J$ is the black hole spin angular momentum, $\tilde{\Sigma} = \Sigma = f$, and
\be
\Sigma = r^2 + a^2 \cos^2\theta \, , \qquad
\Delta = r^2 - 2 M r + a^2 \, , \qquad
B = \left(r^2+a^2\right)A_1-a^2A_2\sin^2\theta \, .
\ee
The functions $f$, $A_1$, $A_2$, and $A_5$ are defined as
\be
f = \sum^\infty_{n=3} \epsilon_n \frac{M^n}{r^{n-2}} \, , \quad
A_1 = 1 + \sum^\infty_{n=3} \alpha_{1n} \left(\frac{M}{r}\right)^n \, , \quad
A_2 = 1 + \sum^\infty_{n=2} \alpha_{2n}\left(\frac{M}{r}\right)^n \, , \quad
A_5 = 1 + \sum^\infty_{n=2} \alpha_{5n}\left(\frac{M}{r}\right)^n \, ,
\ee
where $\{ \epsilon_n \}$, $\{ \alpha_{1n} \}$, $\{ \alpha_{2n} \}$, and $\{ \alpha_{5n} \}$ are four infinite sets of deformation parameters without constraints from the Newtonian limit and Solar System experiments. In this paper, we have only considered the deformation parameters $\epsilon_3$, $\alpha_{13}$, and $\alpha_{22}$ because they are associated to the leading order corrections in $f$, $A_1$, and $A_2$, respectively. We have ignored the leading order correction in $A_5$ because its impact on the reflection spectrum is very weak ~\citep{noim1}. Note that in any model we only consider the possibility that one of the deformation parameters can be non-vanishing and we set all others to zero. For example, when we try to measure $\alpha_{13}$ we leave it free in the fit while all other deformation parameters vanish. The possibility of two (or more) variable deformation parameters at the same time is beyond the capabilities of our current version of {\sc relxill\_nk}~\citep{noim2}.

Note that {\sc relxill\_nk} scans a restricted parameter space of the metric to avoid spacetimes with pathological properties. We require that $| a_* | \le 1$, because for $| a_* | > 1$ there is no black hole but a naked singularity, exactly like in the Kerr metric. As discussed in~\citet{tj,564}, we also have to impose the following restrictions on $\alpha_{13}$, $\alpha_{22}$, and $\epsilon_3$
\be
\label{eq-constraints}
\alpha_{13} > - \frac{1}{2} \left( 1 + \sqrt{1 - a^2_*} \right)^4 \, , \quad
- \left(1 + \sqrt{1 - a_*^2} \right)^2 < \alpha_{22} < \frac{\left( 1 + \sqrt{1 - a^2_*} \right)^4}{a_*^2} \, ,
\quad
\epsilon_3 >  - \left( 1 + \sqrt{1 - a^2_*} \right)^3 \, .
\ee

%%%%%%%%%%%%%%%%%%%%%%%%%%%%%%%

\end{document}